\documentclass[aps,groupedaddress,preprint,superscriptaddress,showpacs]{revtex4-1}
\usepackage{eurosym}
\usepackage{amsmath}
\usepackage{fixmath}
\usepackage{graphicx}
\usepackage{wrapfig}
\usepackage[toc,page]{appendix}
\usepackage{subcaption}

\usepackage{hyperref}
\usepackage[capitalise]{cleveref}
\usepackage{subcaption}

\usepackage{amsfonts}

\graphicspath{{images/}}
\newcommand{\mb}{\mathbold}

\newcommand{\be}{\begin{equation}}
\newcommand{\ee}{\end{equation}}

\begin{document}

\title{Generation and dynamics of entangled fermion-photon-phonon states in
nanocavities}

\author{ Mikhail Tokman}
\affiliation{Institute of Applied Physics, Russian Academy of Sciences, Nizhny Novgorod, 603950, Russia }
\author{ Maria Erukhimova}
\affiliation{Institute of Applied Physics, Russian Academy of Sciences, Nizhny Novgorod, 603950, Russia }
\author{Yongrui Wang}
\affiliation{Department of Physics and Astronomy, Texas A\&M University, College Station,
TX, 77843 USA}
\author{Qianfan Chen}
\affiliation{Department of Physics and Astronomy, Texas A\&M University, College Station,
TX, 77843 USA}
\author{Alexey Belyanin}
\affiliation{Department of Physics and Astronomy, Texas A\&M University, College Station,
TX, 77843 USA}

\begin{abstract}

We develop the analytic theory describing the formation and evolution of entangled quantum states for a fermionic quantum emitter coupled to a quantized electromagnetic field in a nanocavity and quantized phonon or mechanical vibrational modes. The theory is applicable to a broad range of cavity quantum optomechanics problems and  emerging research on plasmonic nanocavities coupled to single molecules and other quantum emitters. The optimal conditions for a tri-state entanglement are realized near the parametric resonances in a coupled system. The model includes decoherence effects due to coupling of the fermion, photon, and phonon subsystems to their dissipative reservoirs within the stochastic evolution approach, which is derived from the Heisenberg-Langevin formalism. Our theory provides analytic expressions for the time evolution of the quantum state and observables, and the emission spectra. The limit of a classical acoustic pumping and the interplay between parametric and standard one-photon resonances are analyzed. 

\end{abstract}

\date{\today }

\maketitle

\section{Introduction}

There is a lot of recent interest in the quantum dynamics of fermion
systems coupled  to an electromagnetic (EM) mode in a cavity and
quantum or classical mechanical/acoustic oscillations or phonon vibrations. This problem is related
to the burgeoning fields of cavity optomechanics \cite{aspelmeyer2014,meystre2013,pirkkalainen2015} and quantum acoustics \cite{chu2017,hong2017, arriola2019}. Another
example where this situation can be realized is a molecule placed in a plasmonic 
nanocavity  \cite{benz2016, park2016}. In this case the fermion
system may comprise two or more electron states forming an optical transition,
whereas the phonon field is simply a vibrational mode of a molecule. One can also imagine a situation where a quantum emitter such as a quantum dot or an optically active defect in a solid matrix is coupled to the quantized phonon modes of a crystal lattice, which would be an extension of an extremely active field of research on phonon-polaritons or plasmon-phonon-polaritons \cite{tame2013,maia2019} into a fully quantum regime. 

Apart from the fundamental interest, the studies of such systems are motivated
by quantum information applications. Indeed, the presence of a classical or
quantized acoustic mode provides an extra handle to control the quantum
state of a coupled fermion-boson quantum system. In the extreme quantum
limit in which the fermionic degree of freedom and all bosonic degrees of
freedom (both photons and phonons) are quantized, a strong enough coupling
between them leads to an entangled fermion-photon-phonon state, which is a
complex enough system to implement basic gates for quantum computation or
other applications. Such a system has not been realized experimentally.
However, many ingredients have been already demonstrated, such as strong
coupling between a nanocavity mode and a single molecule \cite{chikkaraddy2016}, numerous examples of strong coupling between nanocavity modes and a single fermionic quantum emitter such as a color center \cite{lukin2016} or a quantum dot (QD) (see e.g. Refs.~\cite{deppe,reithmaier} for semiconductor cavity-QD systems and Refs. \cite{pelton2018, bitton2019, park2019} for plasmonic cavities),  strong coupling and entanglement of acoustic phonons \cite{satzinger2018, bienfait2020}, resolving the energy levels of a nanomechanical oscillator \cite{arriola2019}, or cooling a macroscopic system
into its motional ground state \cite{delic2020}. 

Interaction of three or more modes of oscillations, whether they
are classical or quantized, is strongly enhanced close to the parametric
resonance, which is therefore the most interesting region to study. Fortunately for theorists, the analysis near the parametric resonance is greatly simplified, because some form of a slowly varying amplitude method for classical
systems \cite{bloem,bogol} or the rotating wave
approximation (RWA) for quantum systems \cite{Scully1997} can be applied.
The use of RWA restricts the coupling strength to the values much lower than
the characteristic energies in the system, such as the optical transition or
vibrational energy. The emerging studies of the so-called ultra-strong
coupling regime \cite{kono2019} have to go beyond the RWA. Nevertheless, for the
vast majority of experiments, including nonperturbative strong coupling
dynamics and entanglement, the RWA is adequate and provides some crucial
simplifications that allow one to obtain analytic solutions.

In particular, within Schr\"{o}dinger's description, the equations of motion
for the components of an infinitely dimensional state vector $\left\vert
\Psi \right\rangle $ that describes a coupled fermion-boson system can be
split into the blocks of low dimensions if the RWA is applied. This is true
even if the dynamics of the fermion subsystem is nonperturbative, e.g. the
effects of saturation are important. Note that there is no such
simplification in the Heisenberg representation, i.e.~when solving the equations
of motion
\begin{equation}
\frac{d}{dt}\hat{g}=\frac{i}{\hbar }\left[ \hat{H},\hat{g}\right] ,
\label{hei}
\end{equation}%
where $\widehat{g}$ is the Heisenberg operator of a certain physical
observable $g$ and $\hat{H}$ is the Hamiltonian of the system.
Operator-valued Eqs.~(\ref{hei}) are generally impossible to split into
smaller blocks, even within the RWA. This happens because some 
matrix elements $g_{AB}(t)$ of  the Heisenberg operator are determined by states $|A\rangle$, $|B\rangle$ which belong to different blocks that evolve independently in the Schr\"{o}dinger picture. The simplification could only be possible for specially selected initial conditions in which the Heisenberg operator is determined on a ``truncated'' basis belonging to only one of the independent blocks.   The Schr\"{o}dinger's approach also leads to fewer equations
for the state vector components than the approach based on the von Neumann
master equation for the elements of the density matrix.

Obviously, the Schr\"{o}dinger equation in its standard form cannot be
applied to describe open systems coupled to a dissipative reservoir. In this
case the stochastic versions of the equation of evolution for the state vector
have been developed, e.g. the method of quantum jumps \cite{Scully1997,Plenio1998}. This method is optimal for numerical analysis in
the Monte-Carlo type schemes. Here we formulate the stochastic equation for
the state vector derived from the Heisenberg-Langevin approach which is more
conducive to the analytic treatment. Its key element is an assumption that
there exists the operator of evolution $\hat{U}$, which is determined
unambiguously not only by the parameters of the dynamical system but also by
the statistical properties of a dissipative reservoir.

The paper is structured as follows. Section II formulates the model and the Hamiltonian for coupled quantized fermion, photon, and phonon fields in a nanocavity. Section III derives the solution for the quantum states of a closed system in the vicinity of a parametric resonance and analyze its properties. In Section IV we provide the stochastic equation describing the evolution of quantum states of an open system in contact with a dissipative reservoir and describe the observables. In Section V we consider the case of a classical acoustic pumping. Section VI describes the interplay of parametric and standard one-photon resonances and provides the conditions under which these resonances can be separated. Section VII gives an example of manipulating entangled electron-photon states by an acoustic pumping. Appendix contains the derivation of the stochastic equation of evolution from the Heisenberg-Langevin approach and compares with Lindblad density-matrix formalism. 

\section{A coupled quantized electron-photon-phonon system: the model}

Consider a quantized electron system coupled to the quantum EM field of a
nanocavity and classical or quantized vibrational (phonon) modes, see Fig.~1 which sketches two out of many possible scenarios. 


\begin{figure}[htb]
\centering
\begin{subfigure}[b]{0.3\textwidth}
\includegraphics[width=1\linewidth]{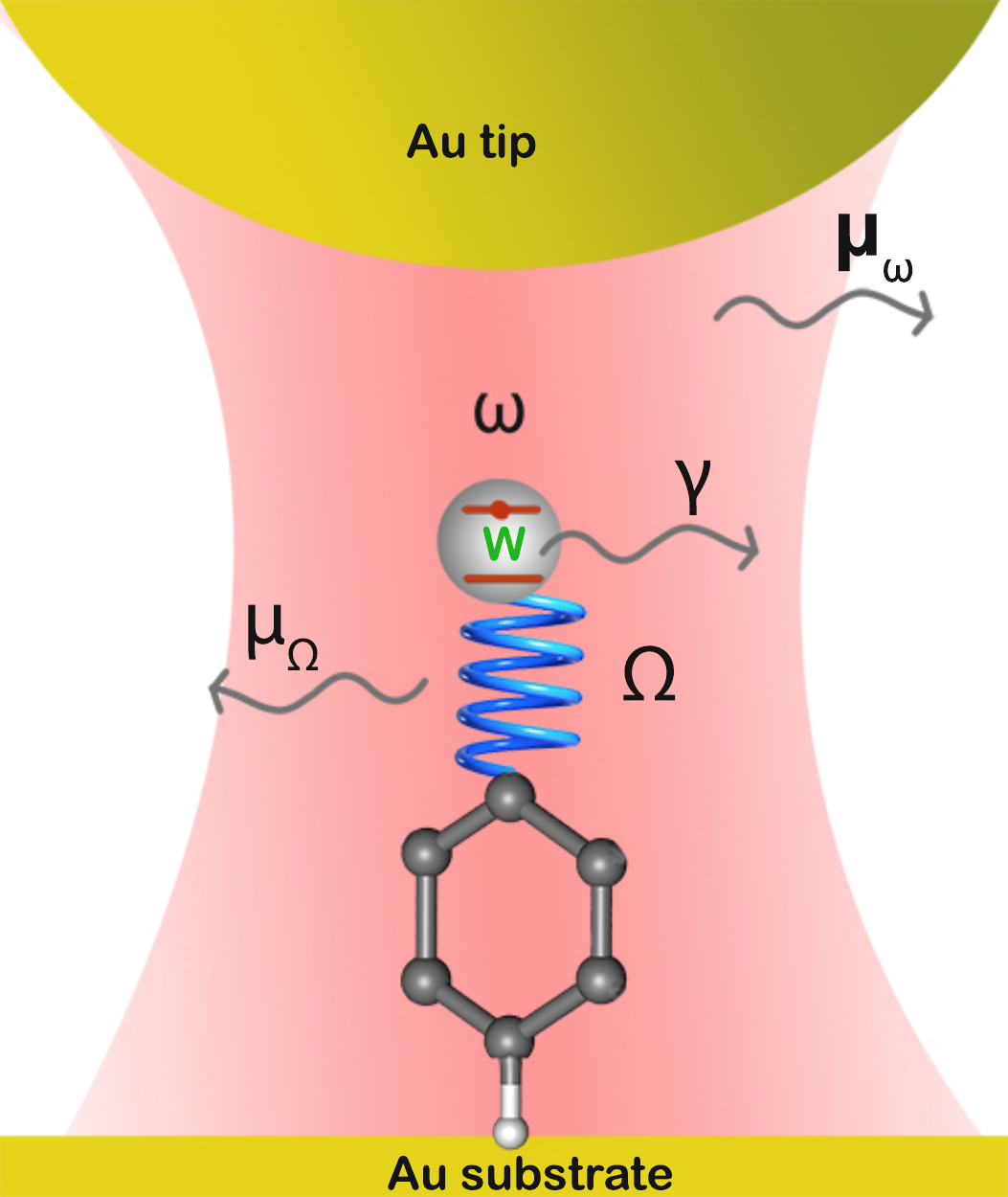}
\caption{ }
\label{fig1a}
\end{subfigure}

\begin{subfigure}[b]{0.3\textwidth}
\includegraphics[width=1\linewidth]{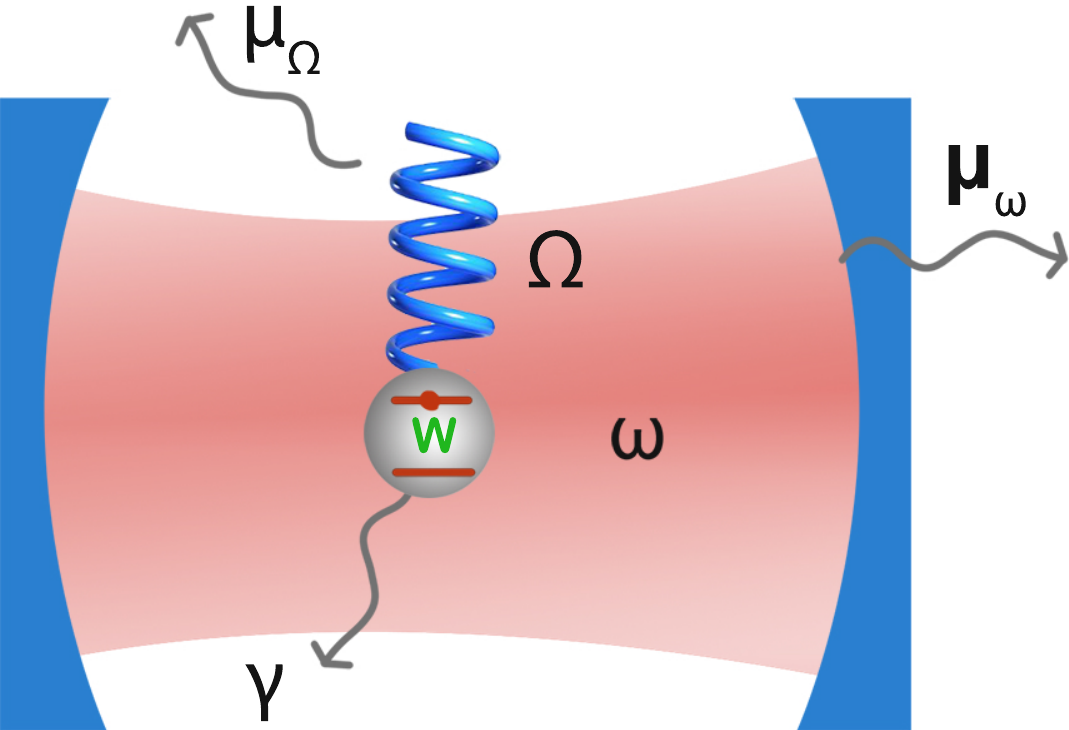}
\caption{  }
\label{fig1b}
\end{subfigure}
\caption{(a) A sketch of a molecule in a nanocavity created by
a metallic nanotip and a substrate;  (b)  A sketch of a quantum dot coupled to optical and mechanical vibrational
modes in a nanocavity.  }
\end{figure}

Here the electron transition energy is $W$, the photon and phonon mode frequencies are $\omega$ and $\Omega$, respectively. The decay constants $\gamma$, $\mu_{\omega}$, and $\mu_{\Omega}$  of the electron, photon, and phonon subsystems due to couplings to their respective dissipative reservoirs are also indicated. Figure 1b implies that it is a quantum dot which experiences vibrations, but our treatment below works for any mechanism of relative displacement between the electron system and the field of an EM cavity mode, including the situations where it is the wall of a nanocavity which experiences oscillations. 

We start from writing down a general Hamiltonian for a coupled quantized
electron-photon-phonon system and derive its various approximate forms: the
RWA, small-amplitude acoustic oscillations, classical vs. quantum phonon
mode, etc.

\subsection{The fermion subsystem}

Consider the simplest version of the fermion subsystem: two electron states $%
\left\vert 0\right\rangle $ and $\left\vert 1\right\rangle $ with energies $%
0 $ and $W$, respectively. We will call it an \textquotedblleft
atom\textquotedblright\ for brevity, although it can be electron states of a molecule, a
quantum dot, or any other electron system. Introduce creation and
annihilation operators of the excited state $\left\vert 1\right\rangle $, 
$
\hat{\sigma}=\left\vert 0\right\rangle \left\langle 1\right\vert$, $\hat{\sigma 
}^{\dagger }=\left\vert 1\right\rangle \left\langle 0\right\vert$,
which satisfy standard commutation relations for fermions:
\begin{equation*}
\hat{\sigma}^{\dagger }\left\vert 0\right\rangle =\left\vert 1\right\rangle ,%
\hat{\sigma}\left\vert 1\right\rangle =\left\vert 0\right\rangle ,\hat{\sigma%
}\hat{\sigma}=\hat{\sigma}^{\dagger }\hat{\sigma}^{\dagger }=0;\left[ \hat{%
\sigma},\hat{\sigma}^{\dagger }\right] _{+}=\hat{\sigma}\hat{\sigma}%
^{\dagger }+\hat{\sigma}^{\dagger }\hat{\sigma}=1.
\end{equation*}%
The Hamiltonian of an atom is%
\begin{equation}
\hat{H}_{a}=W\hat{\sigma}^{\dagger }\hat{\sigma}.  \label{ha}
\end{equation}%
We will also need the dipole moment operator,%
\begin{equation}
\mathbf{\hat{d}}=\mathbf{d}\left( \hat{\sigma}^{\dagger }+\hat{\sigma}%
\right) ,  \label{dip}
\end{equation}%
where $\mathbf{d=}\left\langle 1\right\vert \mathbf{\hat{d}}\left\vert
0\right\rangle $ is a real vector. For a finite motion we can always choose
the coordinate representation of stationary states in terms of real
functions.

\subsection{Quantized EM modes of a cavity}

We use a standard representation for the electric field operator in a cavity:%
\begin{equation}
\mathbf{\hat{E}}=\sum_{i}\left[ \mathbf{E}_{i}\left( \mathbf{r}\right) \hat{c%
}_{i}+\mathbf{E}_{i}^{\ast }\left( \mathbf{r}\right) \hat{c}_{i}^{\dagger }%
\right] ,  \label{ef}
\end{equation}%
where $\hat{c}_{i}^{\dagger },\hat{c}_{i}$ are creation and annihilation
operators for photons at frequency $\omega _{i\text{ }}$; the functions $%
\mathbf{E}_{i}\left( \mathbf{r}\right) $ describe the spatial structure of
the EM modes in a cavity. The relation between the modal frequency $\omega
_{i\text{ }}$ and the function $\mathbf{E}_{i}\left( \mathbf{r}\right) $ can
be found by solving the boundary-value problem of the classical
electrodynamics \cite{Scully1997}. The normalization conditions \cite%
{Tokman2016}%
\begin{equation}
\int_{V}\frac{\partial \left[ \omega_i^{2}\varepsilon \left( \omega_i,\mathbf{r%
}\right) \right] }{\omega_i \partial \omega_i }\mathbf{E}_{i}^{\ast }\left(
\mathbf{r}\right) \mathbf{E}_{i}\left( \mathbf{r}\right) d^{3}r=4\pi \hbar
\omega _{i}  \label{nc}
\end{equation}%
ensure correct bosonic commutators $\left[ \hat{c}_{i},\hat{c}_{i}^{\dagger }%
\right] =\delta _{ij}$ and the field Hamiltonian in the form

\begin{equation}
\hat{H}_{em}=\hbar \sum_{i}\omega _{i}\left( \hat{c}_{i}^{\dagger }\hat{c}%
_{i}+\frac{1}{2}\right) .  \label{hem}
\end{equation}%
Here $V$ is a quantization volume  and $\varepsilon \left( \omega ,\mathbf{r}%
\right) $ is the dielectric function of a dispersive medium that fills the
cavity.

\subsection{The quantized phonon field}

We assume that our two-level atom is dressed by a phonon field
which can be described by the displacement operator:
\begin{equation}
\mathbf{\hat{q}}=\sum_{i}\mathbf{\hat{q}}_{i};~~\mathbf{\hat{q}}_{i}=\mathbf{%
Q}_{i}\left( \mathbf{r}\right) \hat{b}_{i}+\mathbf{Q}_{i}^{\ast }\left(
\mathbf{r}\right) \hat{b}_{i}^{\dagger }  \label{af}
\end{equation}%
Here $\hat{b}_{i}$ and $\hat{b}_{i}^{\dagger }$ are annihilation and
creation operators of phonons and the functions $\mathbf{Q}_{i}\left(
\mathbf{r}\right) $ determine the spatial structure of oscillations at
frequencies $\Omega _{i}$. Expression~(\ref{af}) can be used when the
amplitude of oscillations is small enough. One can always choose the
normalization of functions $\mathbf{Q}_{i}\left( \mathbf{r}\right) $
corresponding to standard commutation relations for bosons, $\left[ \hat{b}%
_{i},\hat{b}_{j}^{\dagger }\right] =\delta _{ij}$ and a standard form for
the Hamiltonian of mechanical oscillations:
\begin{equation}
\hat{H}_{p}=\hbar \sum_{i}\Omega _{i}\left( \hat{b}_{i}^{\dagger }\hat{b}%
_{i}+\frac{1}{2}\right) .  \label{hp}
\end{equation}

\subsection{An atom coupled to quantized EM and phonon fields}

Now we can combine all ingredients into a coupled quantized system. Adding
the interaction Hamiltonian with a EM cavity mode in the electric dipole
approximation, $-\mathbf{\hat{d}}\cdot \mathbf{\hat{E}}$, the Hamiltonian of
an atom coupled to a single mode EM field
\begin{equation}
\hat{H}=\hat{H}_{em}+\hat{H}_{a}-\mathbf{d}\left( \hat{\sigma}^{\dagger }+%
\hat{\sigma}\right) \cdot \left[ \mathbf{E}\left( \mathbf{r}\right) \hat{c}+%
\mathbf{E}^{\ast }\left( \mathbf{r}\right) \hat{c}^{\dagger }\right] _{%
\mathbf{r}=\mathbf{r}_{a}},  \label{h}
\end{equation}%
where $\mathbf{r}=\mathbf{r}_{a}$ denotes the position of an atom inside the
cavity. The effect of \textquotedblleft dressing\textquotedblright\ of the
coupled atom-EM field system by mechanical oscillations in its most general
form can be included by adding the Hamiltonian of phonon modes $\hat{H}%
_{p} $ and substituting $\mathbf{r}_{a}\Longrightarrow \mathbf{r}_{a}+%
\mathbf{\hat{q}}$ in Eq.~(\ref{h}). This will work for an arbitrary relative displacement of an atom with respect to the EM cavity mode. Keeping only one phonon mode for
simplicity, in which%
\begin{equation}
\mathbf{\hat{q}}=\mathbf{Q}\left( \mathbf{r}\right) \hat{b}+\mathbf{Q}^{\ast
}\left( \mathbf{r}\right) \hat{b}^{\dagger },  \label{qo}
\end{equation}%
and expanding in Taylor series in the vicinity of $\mathbf{r}=\mathbf{r}_{a}$%
, we obtain the total Hamiltonian,
\begin{eqnarray}
\hat{H} &=&\hat{H}_{em}+\hat{H}_{a}+\hat{H}_{p}-\left( \chi \hat{\sigma}%
^{\dagger }\hat{c}+\chi ^{\ast }\hat{\sigma}\hat{c}^{\dagger }+\chi \hat{%
\sigma}\hat{c}+\chi ^{\ast }\hat{\sigma}^{\dagger }\hat{c}^{\dagger }\right)
\notag \\
&&-\left( \eta _{1}\hat{\sigma}^{\dagger }\hat{c}\hat{b}+\eta _{1}^{\ast }%
\hat{\sigma}\hat{c}^{\dagger }\hat{b}^{\dagger }+\eta _{2}\hat{\sigma}%
^{\dagger }\hat{c}\hat{b}^{\dagger }+\eta _{2}^{\ast }\hat{\sigma}\hat{c}%
^{\dagger }\hat{b}+\eta _{1}\hat{\sigma}\hat{c}\hat{b}+\eta _{1}^{\ast }\hat{%
\sigma}^{\dagger }\hat{c}^{\dagger }\hat{b}^{\dagger }+\eta _{2}\hat{\sigma}%
\hat{c}\hat{b}^{\dagger }+\eta _{2}^{\ast }\hat{\sigma}^{\dagger }\hat{c}%
^{\dagger }\hat{b}\right)  \label{toh}
\end{eqnarray}%
where%
\begin{equation*}
\chi =\left( \mathbf{\hat{d}}\cdot \mathbf{\hat{E}}\right) _{\mathbf{r}=%
\mathbf{r}_{a}}, \; \eta _{1}=\left[ \mathbf{d}\left( \mathbf{Q\cdot \nabla }%
\right) \mathbf{E}\right] _{\mathbf{r}=\mathbf{r}_{a}}, \; \eta _{2}=\left[
\mathbf{d}\left( \mathbf{Q}^{\ast }\mathbf{\cdot \nabla }\right) \mathbf{E}%
\right] _{\mathbf{r}=\mathbf{r}_{a}}.
\end{equation*}%
Note that we can always take the functions $\mathbf{E}\left( \mathbf{r}%
\right) $ and $\mathbf{Q}\left( \mathbf{r}\right) $ to be real \textit{at
the position of an atom}, but we cannot keep the derivatives real at the
same time if the modal structure $\propto e^{i\mathbf{k\cdot r}}$ . However,
for ideal cavity modes the latter is possible. As we will see below,
the best conditions for electron-photon-phonon entanglement are reached in
the vicinity of the \textit{parametric resonance}:
\begin{equation}
\frac{W}{\hbar }\approx \omega \pm \Omega .  \label{par}
\end{equation}%
When the upper sign is chosen in Eq.~(\ref{par}), the RWA applied to the
Hamiltonian~(\ref{toh}) yields
\begin{equation}
\hat{H}=\hat{H}_{em}+\hat{H}_{a}+\hat{H}_{p}-\left( \eta \hat{\sigma}%
^{\dagger }\hat{c}\hat{b}+\eta ^{\ast }\hat{\sigma}\hat{c}^{\dagger }\hat{b}%
^{\dagger }\right)  \label{upsign}
\end{equation}%
where $\eta \equiv \eta _{1}$. For the lower sign in Eq.~(\ref{par}), the RWA
Hamiltonian is%
\begin{equation}
\hat{H}=\hat{H}_{em}+\hat{H}_{a}+\hat{H}_{p}-\left( \eta \hat{\sigma}%
^{\dagger }\hat{c}\hat{b}^{\dagger }+\eta ^{\ast }\hat{\sigma}\hat{c}%
^{\dagger }\hat{b}\right)  \label{lowersign}
\end{equation}%
where $\eta \equiv \eta _{2}$.

\subsection{An atom coupled to the quantized EM field and dressed by a
classical acoustic field}

For classical acoustic oscillations the operator $\mathbf{\hat{q}}=\mathbf{Q}%
\left( \mathbf{r}\right) \hat{b}+\mathbf{Q}^{\ast }\left( \mathbf{r}\right)
\hat{b}^{\dagger }$ in Eq.~(\ref{qo}) becomes a classical function%
\begin{equation}
\mathbf{q}=\mathbf{Q}\left( \mathbf{r}\right) e^{-i\Omega t}+\mathbf{Q}%
^{\ast }\left( \mathbf{r}\right) e^{i\Omega t}  \label{qc}
\end{equation}%
where $\mathbf{Q}$ is a coordinate-dependent complex amplitude of classical
oscillations. Near the parametric resonance $\left( \omega +\Omega \approx
\frac{W}{\hbar }\right) $ the RWA Hamiltonian takes the form~%
\begin{equation}
\hat{H}=\hat{H}_{em}+\hat{H}_{a}-\left( \mathfrak{R}\hat{\sigma}^{\dagger }\hat{c}%
e^{-i\Omega t}+\mathfrak{R}^{\ast }\hat{\sigma}\hat{c}^{\dagger }e^{i\Omega t}\right) .
\label{sch}
\end{equation}%
where $\mathfrak{R}=\left[ \mathbf{d}\left( \mathbf{Q\cdot \nabla }\right) \mathbf{E}%
\right] _{\mathbf{r}=\mathbf{r}_{a}}$ . The value of the acoustic frequency $%
\Omega $ in Eq.~(\ref{sch}) can be of either sign, corresponding to the
choice \textquotedblleft $\pm $\textquotedblright\ in the parametric
resonance condition Eq.~(\ref{par}); When the sign of $\Omega $ changes from
positive to negative, one should replace $\mathbf{Q}$ with $\mathbf{Q}^{\ast
}$ in the above expression for $\mathfrak{R}$.

\qquad Qualitatively, Hamiltonian~(\ref{upsign}) corresponds to the decay of
the fermionic excitation into a photon and phonon; Hamiltonian~(\ref%
{lowersign}) corresponds to the decay of a photon into a phonon and
fermionic excitation, whereas Hamiltonian~(\ref{sch}) describes parametric
decay of a photon into an atomic excitation and back, mediated by classical
acoustic oscillations.

\section{Parametric resonance in a closed system}

When the system is closed and there is no dissipation, the general analytic
solution to the dynamics of coupled fermions, photons, and phonons can be
obtained in the RWA. We write the state vector as

\begin{equation}
\Psi =\sum_{\alpha ,n=0}^{\infty }\left( C_{\alpha n0}\left\vert \alpha
\right\rangle \left\vert n\right\rangle \left\vert 0\right\rangle +C_{\alpha
n1}\left\vert \alpha \right\rangle \left\vert n\right\rangle \left\vert
1\right\rangle \right) .  \label{psi}
\end{equation}%
Here Greek letters denote phonon states, Latin letters denote photon states,
and numbers $0$, $1$ describe fermion states. We will keep the same sequence
of symbols throughout the paper:
\begin{equation*}
C_{\rm phonon\,photon\, fermion}\left\vert phonon\right\rangle \left\vert
photon\right\rangle \left\vert fermion\right\rangle .
\end{equation*}

Next, we substitute Eq.~(\ref{psi}) into the Schr\"{o}dinger equation,
\begin{equation}
i\hbar \frac{\partial }{\partial t}\left\vert \Psi \right\rangle =\hat{H}%
\left\vert \Psi \right\rangle  \label{sce}
\end{equation}%
Where $\hat{H}$ is the RWA Hamiltonian. For definiteness, we consider the
vicinity of the parametric resonance with a plus sign, $\omega +\Omega
\approx \frac{W}{\hbar }$, which corresponds to the Hamiltonian (\ref%
{upsign}). In this case the equations for the coefficients in Eq.~(\ref{psi}%
) can be separated into the pairs of coupled equations
\begin{equation}
\frac{d}{dt}\left(
\begin{array}{c}
C_{\alpha n0} \\
C_{\left( \alpha -1\right) \left( n-1\right) 1}%
\end{array}%
\right) +\left(
\begin{array}{cc}
i\omega _{\alpha ,n} & -i\Omega _{R}^{\left( \alpha ,n\right) \ast } \\
-i\Omega _{R}^{\left( \alpha ,n\right) } & i\omega _{\alpha ,n}-i\Delta%
\end{array}%
\right) \left(
\begin{array}{c}
C_{\alpha n0} \\
C_{\left( \alpha -1\right) \left( n-1\right) 1}%
\end{array}%
\right) =0,  \label{coeff eq for psi}
\end{equation}%
and a separate equation for the lowest-energy state:%
\begin{equation}
\overset{\cdot }{C}_{000}+i\omega _{0,0}C_{000}=0,  \label{eq for les}
\end{equation}%
where%
\begin{equation*}
\Omega _{R}^{\left( \alpha ,n\right) }=\frac{\eta }{\hbar }\sqrt{\alpha n}
, \; \omega _{\alpha ,n}=\Omega \left( \alpha +\frac{1}{2}\right) +\omega \left(
n+\frac{1}{2}\right) , \; \Delta =\Omega +\omega -\frac{W}{\hbar }.
\end{equation*}%
Note that approximate Eqs.~(\ref{coeff eq for psi}),(\ref{eq for
les}) preserve the norm \textit{exactly}:
\begin{equation*}
\left\vert C_{000}\right\vert ^{2}+\sum_{\alpha =1,n=1}^{\infty ,\infty
}\left( \left\vert C_{\alpha n0}\right\vert ^{2}+\left\vert C_{\left( \alpha
-1\right) \left( n-1\right) 1}\right\vert ^{2}\right) =\sum_{\alpha
=0,n=0}^{\infty ,\infty }\left( \left\vert C_{\alpha n0}\right\vert
^{2}+\left\vert C_{\alpha n1}\right\vert ^{2}\right) = {\rm const}.
\end{equation*}%
The solution to Eq.~(\ref{eq for les}) is trivial: $C_{000}\left( t\right)
=C_{000}\left( 0\right) \exp \left( -i\omega _{0,0}t\right) $. 
The solution to Eqs.~(\ref{coeff eq for psi}) takes the form%
\begin{equation}
\left(
\begin{array}{c}
C_{\alpha n0} \\
C_{\left( \alpha -1\right) \left( n-1\right) 1}%
\end{array}%
\right) =Ae^{-\Lambda _{1}^{\left( \alpha ,n\right) }t}\left(
\begin{array}{c}
1 \\
a_{1}^{\left( \alpha ,n\right) }%
\end{array}%
\right) +Be^{-\Lambda _{2}^{\left( \alpha ,n\right) }t}\left(
\begin{array}{c}
1 \\
a_{2}^{\left( \alpha ,n\right) }%
\end{array}%
\right) ,  \label{sol to coeff eq for psi}
\end{equation}%
where the constants $A$ and $B$ are determined from initial conditions.
Here the eigenvalues $\Lambda
_{1,2}^{\left( \alpha ,n\right) }$ and eigenvectors $\left(
\begin{array}{c}
1 \\
a_{1,2}^{\left( \alpha ,n\right) }%
\end{array}%
\right) $ of the matrix of coefficients in Eq.~(\ref{coeff eq for psi}) are given by 
\begin{equation}
\Lambda _{1,2}^{\left( \alpha ,n\right) }=i\omega _{\alpha ,n}-i\delta
_{1,2}^{\left( \alpha ,n\right) },a_{1,2}^{\left( \alpha ,n\right) }=\frac{%
\delta _{1,2}^{\left( \alpha ,n\right) }}{\Omega _{R}^{\left( \alpha
,n\right) \ast }},  \label{eigenva and eigenve of moc}
\end{equation}%
where%
\begin{equation}
\delta _{1,2}^{\left( \alpha ,n\right) }=\frac{\Delta }{2}\pm \sqrt{\frac{%
\Delta ^{2}}{4}+\left\vert \Omega _{R}^{\left( \alpha ,n\right) }\right\vert
^{2}}.  \label{delta}
\end{equation}%


\begin{figure}[htb]
\begin{center}
\includegraphics[scale=0.4]{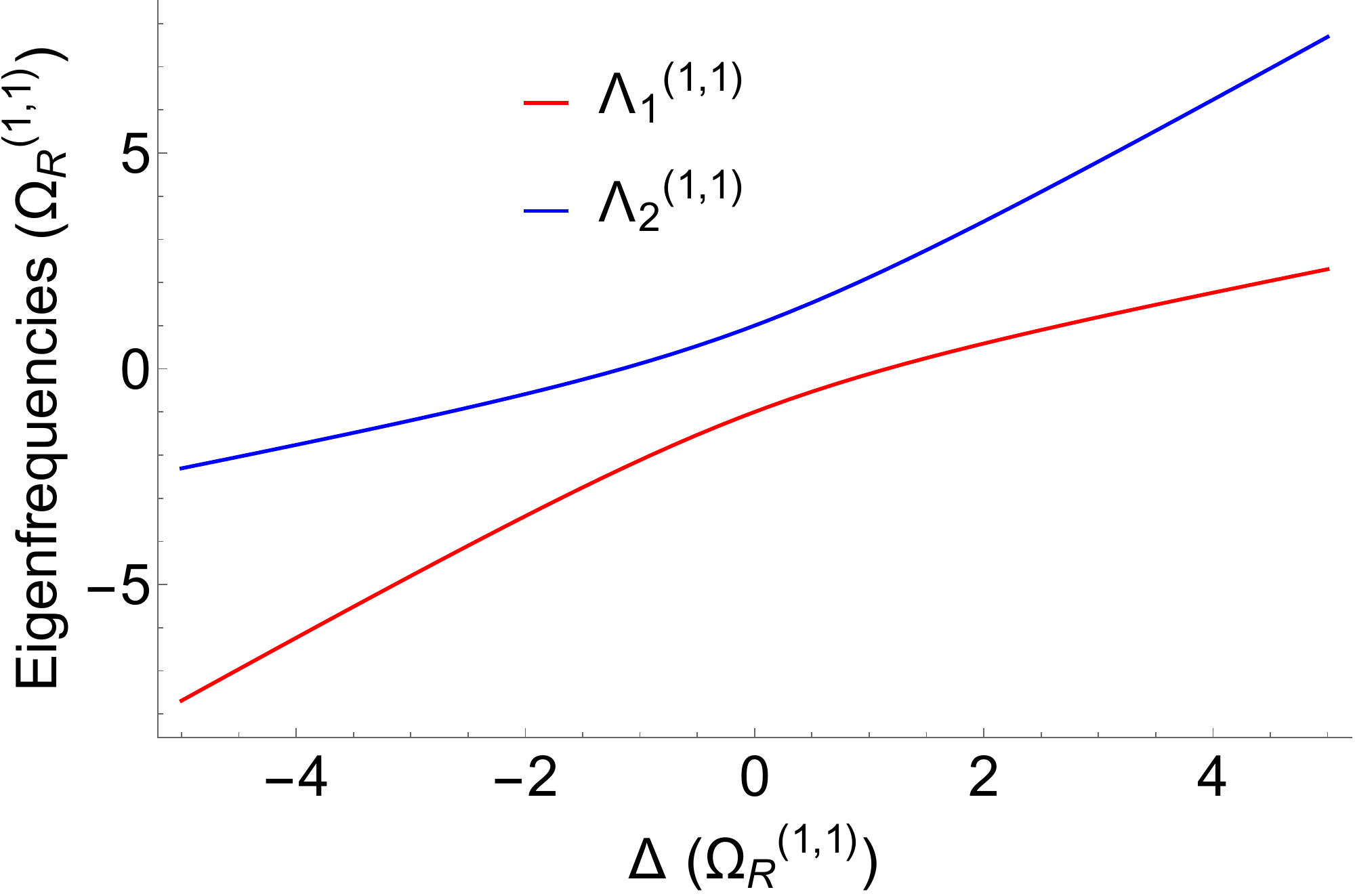}
\caption{Frequency eigenvalues of the coupled electron-photon-phonon quantum system as a function of detuning from the parametric resonance $\frac{W}{\hbar }
=\Omega +$ $\omega $. All frequencies are in units of the generalized Rabi frequency $\Omega _{R}^{\left(1,1\right) }$. The values of eigenfrequencies are shifted vertically by  $\omega_{1,1}|_{\Delta=0}$.     }
\label{fig2}
\end{center}
\end{figure}

Fig. 2 shows the eigenfrequencies of the system given by Eq.~(\ref{eigenva and eigenve of moc}) with $\alpha = n = 1$, shifted by  $\omega_{1,1}|_{\Delta=0}$. One can see the anticrossing with splitting by $2\Omega _{R}^{\left(1,1\right) }$ at the parametric resonance.

As an example, consider an exact parametric resonance, $\frac{W}{\hbar }%
=\Omega +$ $\omega $ and the simplest initial state $\Psi _{0}=\left\vert
0\right\rangle \left\vert 0\right\rangle \left\vert 1\right\rangle $ corresponding to the initially excited atom in a cavity. In
this case the only nonzero amplitudes are $C_{001}$ and $C_{110}$:%
\begin{equation}
\left(
\begin{array}{c}
C_{110} \\
C_{001}%
\end{array}%
\right) =\frac{1}{2}e^{-i\left( \omega _{1,1}-\left\vert \Omega _{R}^{\left(
1,1\right) }\right\vert \right) t}\left(
\begin{array}{c}
e^{-i\theta } \\
1 
\end{array}%
\right) +\frac{1}{2}e^{-i\left( \omega _{1,1}+\left\vert \Omega _{R}^{\left(
1,1\right) }\right\vert \right) t}\left(
\begin{array}{c}
- e^{-i\theta }  \\
1%
\end{array}%
\right) ,  \label{c110 and c001}
\end{equation}%
where%
\begin{equation*}
\omega _{1,1}=\Omega \left( 1+\frac{1}{2}\right) +\omega \left( 1+\frac{1}{2}%
\right) ,\Omega _{R}^{\left( 1,1\right) }=\frac{\eta }{\hbar }=\left\vert
\Omega _{R}^{\left( 1,1\right) }\right\vert e^{i\theta }.
\end{equation*}%
The resulting state vector is%
\begin{equation}
\Psi =e^{-i\omega _{1,1}t}\left[ i e^{-i\theta } \sin \left( \left\vert \Omega _{R}^{\left(
1,1\right) }\right\vert t\right) \left\vert 1\right\rangle \left\vert
1\right\rangle \left\vert 0\right\rangle +\cos \left( \left\vert
\Omega _{R}^{\left( 1,1\right) }\right\vert t\right) \left\vert
0\right\rangle \left\vert 0\right\rangle \left\vert 1\right\rangle \right] .
\label{rsv}
\end{equation}%
This is clearly an entangled electron-photon-phonon state, which is not
surprising. In the absence of dissipation, any coupling between these
subsystems leads to entanglement.

The dynamics of the corresponding physical observables, such as the energy of the field and the atom, is Rabi oscillations
at the frequency which generalizes a standard Rabi frequency to the case of a parametric photon-phonon-atom resonance and which depends on both the spatial structure of the photon and
phonon fields and their occupation numbers:%
\begin{equation}
\left\langle \Psi \right\vert \mathbf{\hat{E}}^{2}\left\vert \Psi
\right\rangle =\left\vert \mathbf{E}\left( \mathbf{r}\right) \right\vert^{2}%
\left[ 2-\cos \left( 2\left\vert \Omega _{R}^{\left( 1,1\right) }\right\vert
t\right) \right]  \label{av of e square}
\end{equation}%
\begin{equation}
\left\langle \Psi \right\vert \hat{H}_{a}\left\vert \Psi \right\rangle =W%
\frac{1+\cos \left( 2\left\vert \Omega _{R}^{\left( 1,1\right) }\right\vert
t\right) }{2}  \label{av of ha}
\end{equation}

\begin{figure}[htb]
\centering
\begin{subfigure}[b]{0.5\textwidth}
\includegraphics[width=1\linewidth]{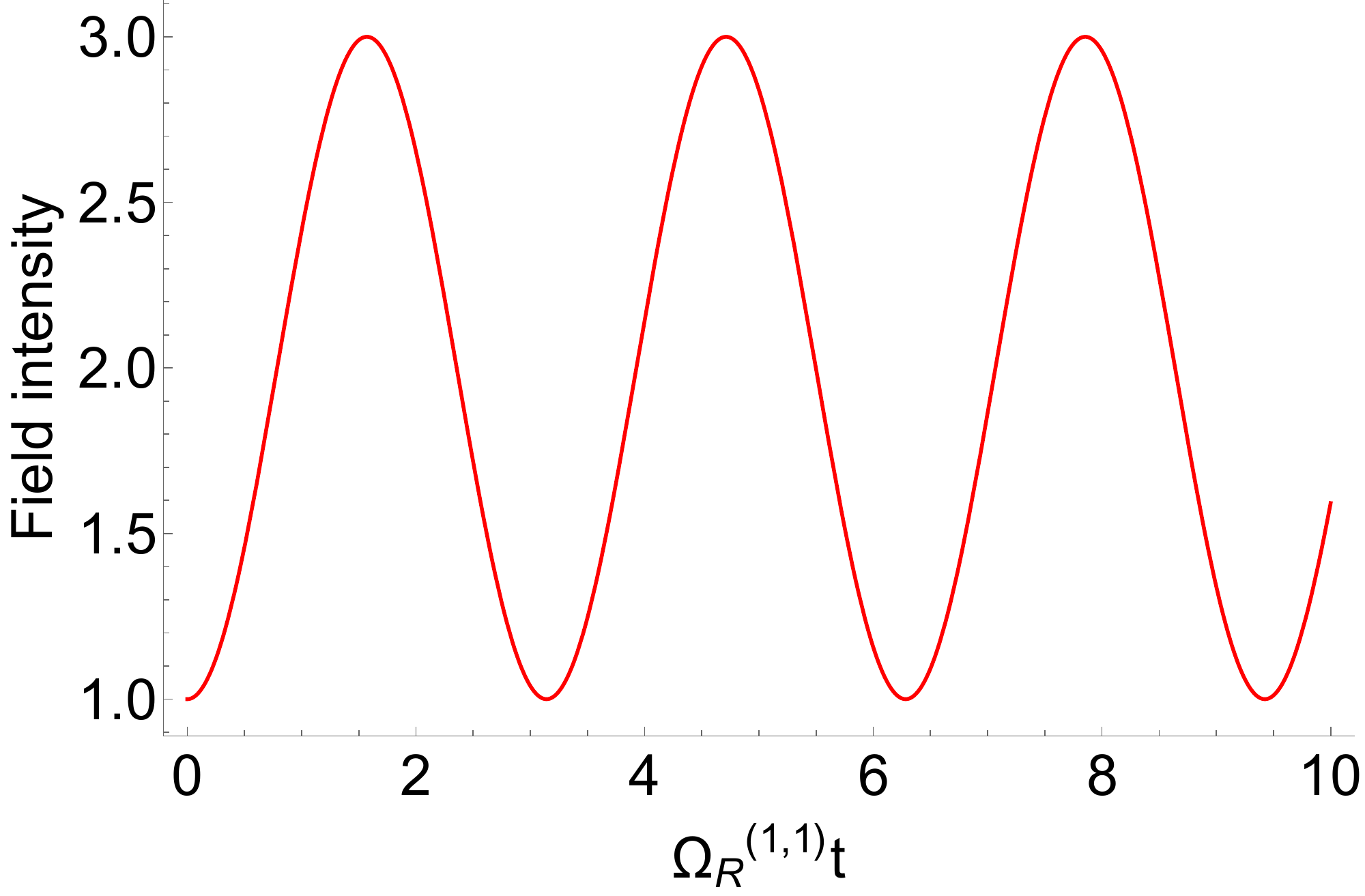}
\caption{ }
\label{fig3a}
\end{subfigure}

\begin{subfigure}[b]{0.5\textwidth}
\includegraphics[width=1\linewidth]{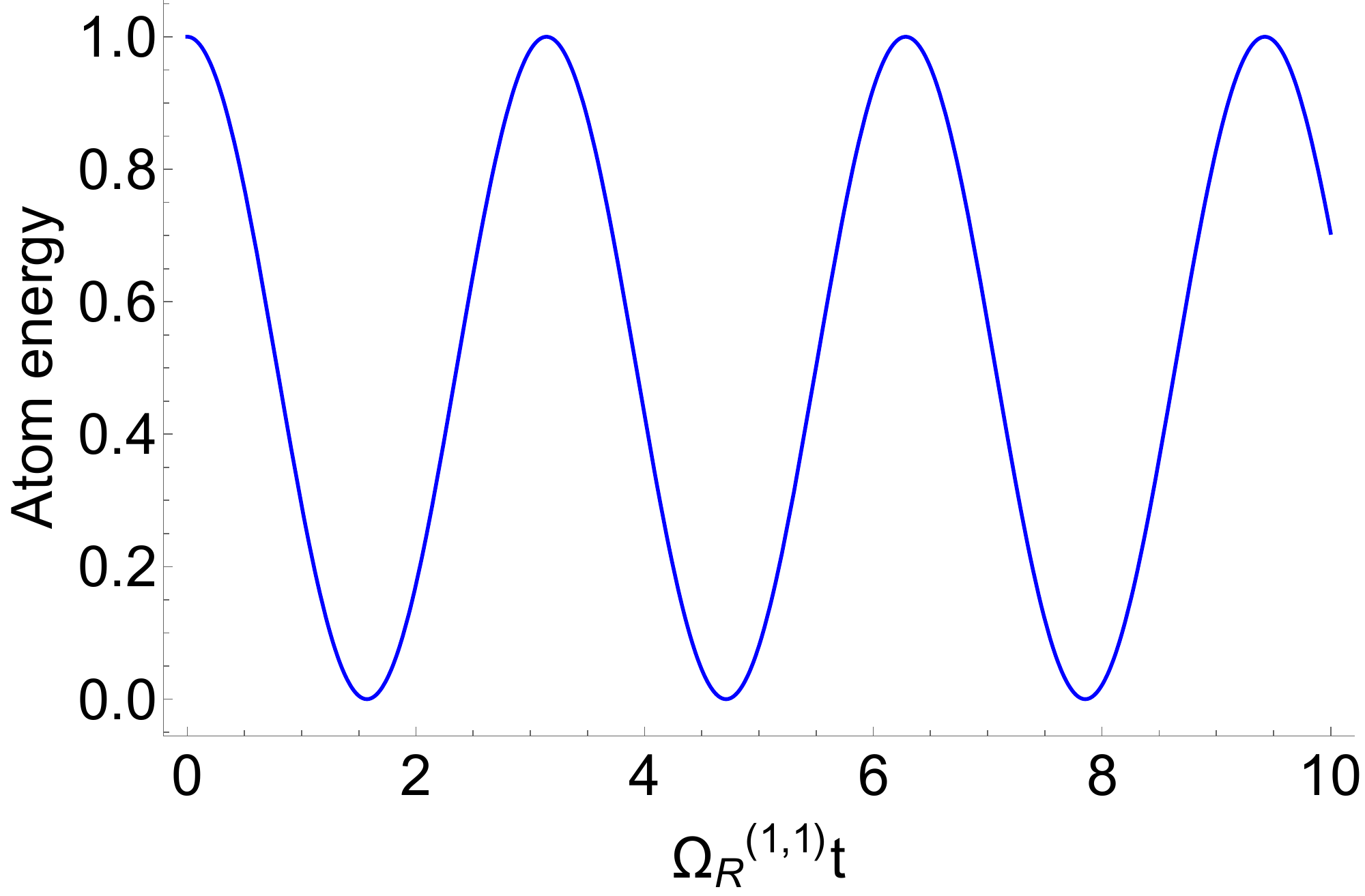}
\caption{  }
\label{fig3b}
\end{subfigure}
\caption{ (a) Normalized field intensity, $\left\langle \Psi \right\vert \mathbf{\hat{E}}^{2} \left\vert \Psi
\right\rangle / \left\vert \mathbf{E}\left( \mathbf{r}\right) \right\vert^{2}$, and  (b)  normalized atom energy $\left\langle \Psi \right\vert \hat{H}_{a}\left\vert \Psi \right\rangle /W$ as a function of time in units of the generalized Rabi frequency $\Omega _{R}^{\left( 1,1\right)}$.   }
\end{figure}

It is illustrated in Fig. 3 which shows the normalized EM field energy density and energy of an atom as a function of time. Note that the EM field energy never reaches zero because of the presence of zero-point vacuum energy. With detuning from the parametric resonance, the amplitude of the oscillations will decrease.

\section{Dynamics of an open electron-photon-phonon system}

\subsection{Stochastic evolution equation}

Now we include the processes of relaxation and decoherence in an open
system, which is (weakly) coupled to a dissipative reservoir. We will use
the stochastic equation of evolution for the state vector, which is derived
in Appendix. This is basically the Schr\"{o}dinger equation modified by
adding a linear relaxation operator and the noise source term with
appropriate correlation properties. The latter are related to the parameters
of the relaxation operator, which is a manifestation of the
fluctuation-dissipation theorem \cite{Landau1965}. In Appendix we derived
the general form of the stochastic equation of evolution from the
Heisenberg-Langevin equations \cite{Scully1997, Gardiner2004, Tokman2013}
and showed how physically reasonable constraints on the observables
determine the properties of the noise sources. We also demonstrated the
relationship between our approach and the Lindblad method of solving the
master equation.

Within our approach the system is described by a state vector which has a
fluctuating component: $\left\vert \Psi \right\rangle =\overline{\left\vert
\Psi \right\rangle }+\widetilde{\left\vert \Psi \right\rangle }$ , where the
straight bar means averaging over the statistics of noise and the wavy bar
denotes the fluctuating component. This state vector is of course very
different from the state vector obtained by solving a standard Schr\"{o}%
dinger equation for a closed system. In fact, coupling to a dissipative
reservoir leads to the formation of a mixed state, which can be described by
a density matrix $\hat{\rho}=\overline{\left\vert \Psi \right\rangle }\cdot
\overline{\left\langle \Psi \right\vert }+\overline{\widetilde{\left\vert
\Psi \right\rangle }\widetilde{\left\langle \Psi \right\vert }}$ . One
should view the stochastic equation approach as a convenient formalism for
calculating physical observables.

Following the derivation in Appendix, Eqs.~(\ref{coeff eq for psi}),(\ref%
{eq for les}) are modified as

\begin{eqnarray}
&&\frac{d}{dt}\left(
\begin{array}{c}
C_{\alpha n0} \\
C_{\left( \alpha -1\right) \left( n-1\right) 1}%
\end{array}%
\right) +\left(
\begin{array}{cc}
i\omega _{\alpha ,n}+\gamma _{\alpha n0} & -i\Omega _{R}^{\left( \alpha
,n\right) \ast } \\
-i\Omega _{R}^{\left( \alpha ,n\right) } & i\omega _{\alpha ,n}-i\Delta
+\gamma _{\left( \alpha -1\right) \left( n-1\right) 1}%
\end{array}%
\right) \left(
\begin{array}{c}
C_{\alpha n0} \\
C_{\left( \alpha -1\right) \left( n-1\right) 1}%
\end{array}%
\right)  \notag \\
&=&-\frac{i}{\hbar }\left(
\begin{array}{c}
R_{\alpha n0} \\
R_{\left( \alpha -1\right) \left( n-1\right) 1}%
\end{array}%
\right) ,  \label{modify coeff eq for psi}
\end{eqnarray}%
\begin{equation}
\dot{C}_{000}+\left( i\omega _{0,0}+\gamma _{000}\right) C_{000}=-\frac{i}{%
\hbar }R_{000}.  \label{modify eq for les}
\end{equation}%
Coupling to a reservoir introduces two main differences to Eqs.~(\ref{modify coeff
eq for psi}),(\ref{modify eq for les}) as compared to Eqs.~(\ref{coeff eq
for psi}),(\ref{eq for les}) for a closed system. First, eigenfrequencies
acquire imaginary parts which describe relaxation:
\begin{equation*}
\omega _{\alpha ,n}\Longrightarrow \omega _{\alpha ,n}-i\gamma _{\alpha
n0}, \, \omega _{\alpha ,n}-\Delta \Longrightarrow \omega _{\alpha ,n}-\Delta
-i\gamma _{\left( \alpha -1\right) \left( n-1\right) 1}, \, \omega
_{0,0}\Longrightarrow \omega _{0,0}-i\gamma _{000}.
\end{equation*}%
The relaxation constants are determined by the properties of all subsystems.
They are derived in Appendix and their explicit form is given in the end
of this section.

Second, the right-hand side of Eqs.~(\ref{modify coeff eq for psi}) and (\ref%
{modify eq for les}) contain noise sources $-\frac{i}{\hbar }R_{\alpha n0}$, 
$-\frac{i}{\hbar }R_{\left( \alpha -1\right) \left( n-1\right) 1}$ and $-%
\frac{i}{\hbar }R_{000}$. They are equal to $0$ after averaging over the
noise statistics: $\overline{R_{\alpha n0}}=\overline{R_{\left( \alpha
-1\right) \left( n-1\right) 1}}=\overline{R_{000}}$. The averages of the
quadratic combinations of noise source terms are nonzero. Including the
noise sources is crucial for consistency of the formalism: it ensures the
conservation of the norm of the state vector and leads to a physically
meaningful equilibrium state. Note that the Weisskopf-Wigner theory does not enforce the conservation of 
 the norm.

\subsection{Evolution of the state amplitudes and observables} 

The solution to Eq.~(\ref{modify eq for les}) is%
\begin{equation}
C_{000}=e^{-\left( i\omega _{0,0}+\gamma _{000}\right) t}\left[
C_{000}\left( 0\right) -\frac{i}{\hbar }\int_{0}^{t}e^{\left( i\omega
_{0,0}+\gamma _{000}\right) \tau }R_{000}\left( \tau \right) d\tau \right] .
\label{modify c000}
\end{equation}%
The solution to Eqs.~(\ref{modify coeff eq for psi}) is determined again by
the eigenvalues and eigenvectors of the matrix of coefficients, 
which are now modified by relaxation rates:
\begin{equation}
\Lambda _{1,2}^{\left( \alpha ,n\right) }=i\omega _{\alpha ,n}-i\delta
_{1,2}^{\left( \alpha ,n\right) },a_{1,2}^{\left( \alpha ,n\right) }=\frac{%
\delta _{1,2}^{\left( \alpha ,n\right) }-i\gamma _{\alpha n0}}{\Omega
_{R}^{\left( \alpha ,n\right) \ast }},
\label{modify eigenva and eigenve of moc}
\end{equation}%
where%
\begin{equation}
\delta _{1,2}^{\left( \alpha ,n\right) }=\frac{\Delta }{2}+i\frac{\gamma
_{\alpha n0}+\gamma _{\left( \alpha -1\right) \left( n-1\right) 1}}{2}\pm
\sqrt{\frac{\left[ \Delta +i\left( \gamma _{\left( \alpha -1\right) \left(
n-1\right) 1}-\gamma _{\alpha n0}\right) \right] ^{2}}{4}+\left\vert \Omega
_{R}^{\left( \alpha ,n\right) }\right\vert ^{2}}.
\label{deltas}
\end{equation}

The solution to Eqs.~(\ref{modify coeff eq for psi}) takes the form%
\begin{eqnarray}
&&\left(
\begin{array}{c}
C_{\alpha n0} \\
C_{\left( \alpha -1\right) \left( n-1\right) 1}%
\end{array}%
\right)  \notag \\
&=&e^{-\Lambda _{1}^{\left( \alpha ,n\right) }t}\left(
\begin{array}{c}
1 \\
a_{1}^{\left( \alpha ,n\right) }%
\end{array}%
\right) \left( A-\frac{i}{\hbar }\int_{0}^{t}e^{\Lambda _{1}^{\left( \alpha
,n\right) }\tau }\frac{R_{\alpha n0}\left( \tau \right) a_{2}^{\left( \alpha
,n\right) }-R_{\left( \alpha -1\right) \left( n-1\right) 1}\left( \tau
\right) }{a_{2}^{\left( \alpha ,n\right) }-a_{1}^{\left( \alpha ,n\right) }}%
d\tau \right)  \notag \\
&&+e^{-\Lambda _{2}^{\left( \alpha ,n\right) }t}\left(
\begin{array}{c}
1 \\
a_{2}^{\left( \alpha ,n\right) }%
\end{array}%
\right) \left( B-\frac{i}{\hbar }\int_{0}^{t}e^{\Lambda _{2}^{\left( \alpha
,n\right) }\tau }\frac{R_{\left( \alpha -1\right) \left( n-1\right) 1}\left(
\tau \right) -R_{\alpha n0}\left( \tau \right) a_{1}^{\left( \alpha
,n\right) }}{a_{2}^{\left( \alpha ,n\right) }-a_{1}^{\left( \alpha ,n\right)
}}d\tau \right)  \label{sol to modify coeff eq for psi}
\end{eqnarray}%
Where the constants $A$ and $B$ are determined by initial conditions.

As an example, we consider the reservoir at low temperatures, when the
steady-state population should go to the ground state $\left\vert
0\right\rangle \left\vert 0\right\rangle \left\vert 0\right\rangle $ . In
this case we can take $\gamma _{000}=0$, as shown below. We will also assume
that the only nonzero correlator of noise is delta-correlated in time:
\begin{equation}
\overline{R_{000}\left( t+\xi \right) R_{000}^{\ast }\left( t\right) }=\hbar
^{2}\delta \left( \xi \right) D_{000}.  \label{con}
\end{equation}%
Then Eqs.~(\ref{modify eq for les}) and~(\ref{modify c000}) yield%
\begin{equation}
\frac{d}{dt}\overline{\left\vert C_{000}\right\vert ^{2}}=D_{000},
\label{eq for av of c000}
\end{equation}%
whereas Eqs.~(\ref{modify coeff eq for psi}) give%
\begin{equation}
\frac{d}{dt}\left( \overline{\left\vert C_{\alpha n0}\right\vert ^{2}}+%
\overline{\left\vert C_{\left( \alpha -1\right) \left( n-1\right)
1}\right\vert ^{2}}\right) =-2\left( \gamma _{\alpha n0}\overline{\left\vert
C_{\alpha n0}\right\vert ^{2}}+\gamma _{\left( \alpha -1\right) \left(
n-1\right) 1}\overline{\left\vert C_{\left( \alpha -1\right) \left(
n-1\right) 1}\right\vert ^{2}}\right) .  \label{eq for av of calphan0}
\end{equation}%
This equation guarantees that the system occupies the ground state at $%
t\rightarrow \infty $.

The noise intensity is determined by the condition that the norm of the
state vector be conserved. This gives
\begin{equation}
D_{000}=2\sum_{\alpha =1,n=1}^{\infty ,\infty }\left( \gamma _{\alpha n0}%
\overline{\left\vert C_{\alpha n0}\right\vert ^{2}}+\gamma _{\left( \alpha
-1\right) \left( n-1\right) 1}\overline{\left\vert C_{\left( \alpha
-1\right) \left( n-1\right) 1}\right\vert ^{2}}\right) .  \label{d000}
\end{equation}%
In Appendix we discuss in detail the dependence of the noise correlator on
the averaged dyadic components of the state vector. We also show how to find
the correlators which ensure that the system approaches thermal distribution
at a finite temperature.

The above formalism allows us to obtain analytic solutions to the state
vector and observables at any temperatures and detunings from the parametric
resonance, while still within the RWA limits. However, the resulting
expressions are very cumbersome and they are better to visualize in the
plots. Let's give an example of the solution at zero reservoir temperature
and exactly at the parametric resonance $\frac{W}{\hbar }=\Omega +$ $\omega $
, when the expressions are more manageable. Consider the initial state $\Psi
_{0}=\left\vert 0\right\rangle \left\vert 0\right\rangle \left\vert
1\right\rangle $ when an atom is excited and boson modes are in the ground
state. In this case the only nonzero amplitudes are $C_{000}$, $C_{001}$ and
$C_{110}$. To make the algebra a bit simpler, we assume that the dissipation
is weak enough and its effect on the eigenvectors $\left(
\begin{array}{c}
1 \\
a_{1,2}^{\left( \alpha ,n\right) }%
\end{array}%
\right) $ can be neglected. As a result, we obtain%
\begin{equation}
\Psi =e^{-\left( i\omega _{1,1}-\frac{\gamma_{001}+\gamma_{110}}{2}%
\right) t}\left[ i e^{-i\theta } \sin \left( \left\vert \widetilde{\Omega }_{R}^{\left(
1,1\right) }\right\vert t\right) \left\vert 1\right\rangle \left\vert
1\right\rangle \left\vert 0\right\rangle +\cos \left( \left\vert
\widetilde{\Omega }_{R}^{\left( 1,1\right) }\right\vert t\right) \left\vert
0\right\rangle \left\vert 0\right\rangle \left\vert 1\right\rangle \right]
+C_{000}\left\vert 0\right\rangle \left\vert 0\right\rangle \left\vert
0\right\rangle ,  \label{modify rsv}
\end{equation}%
where%
\begin{equation*}
\overline{\left\vert C_{000}\right\vert ^{2}}=1-e^{-\left( \gamma_{110}+\gamma _{001}\right) t}, \; \widetilde{\Omega }_{R}^{\left( 1,1\right)
}=\sqrt{\left\vert \Omega _{R}^{\left( 1,1\right) }\right\vert ^{2}-\frac{%
\left( \gamma _{001}-\gamma_{110}\right) ^{2}}{4}}, \; \theta ={\rm Arg}\left[
\Omega _{R}^{\left( 1,1\right) }\right] .
\end{equation*}%
As we see, dissipation leads not only to the relaxation of the entangled
part of the state vector, but also to the frequency shift of the Rabi
oscillations. This shift is absent if $\gamma _{001}=\gamma_{110}$.

The resulting expressions for the observables, such as the EM field
intensity and the energy of the atomic excitation are
\begin{equation}
\left\langle \Psi \right\vert \mathbf{\hat{E}}^{2}\left\vert \Psi
\right\rangle =\left\vert \mathbf{E}\left( \mathbf{r}\right) \right\vert ^{2}%
\left[ 1+e^{-\left( \gamma_{110}+\gamma _{001}\right) t}-\cos \left(
2\left\vert \widetilde{\Omega }_{R}^{\left( 1,1\right) }\right\vert t\right)
e^{-\left( \gamma_{110}+\gamma _{001}\right) t}\right] ,
\label{modify av of e square}
\end{equation}%
\begin{equation}
\left\langle \Psi \right\vert \hat{H}_{a}\left\vert \Psi \right\rangle =W%
\frac{1+\cos \left( 2\left\vert \widetilde{\Omega }_{R}^{\left( 1,1\right)
}\right\vert t\right) }{2}e^{-\left( \gamma _{110}+\gamma _{001}\right) t}
\label{modify av of  ha}
\end{equation}


\begin{figure}[htb]
\centering
\begin{subfigure}[b]{0.5\textwidth}
\includegraphics[width=1\linewidth]{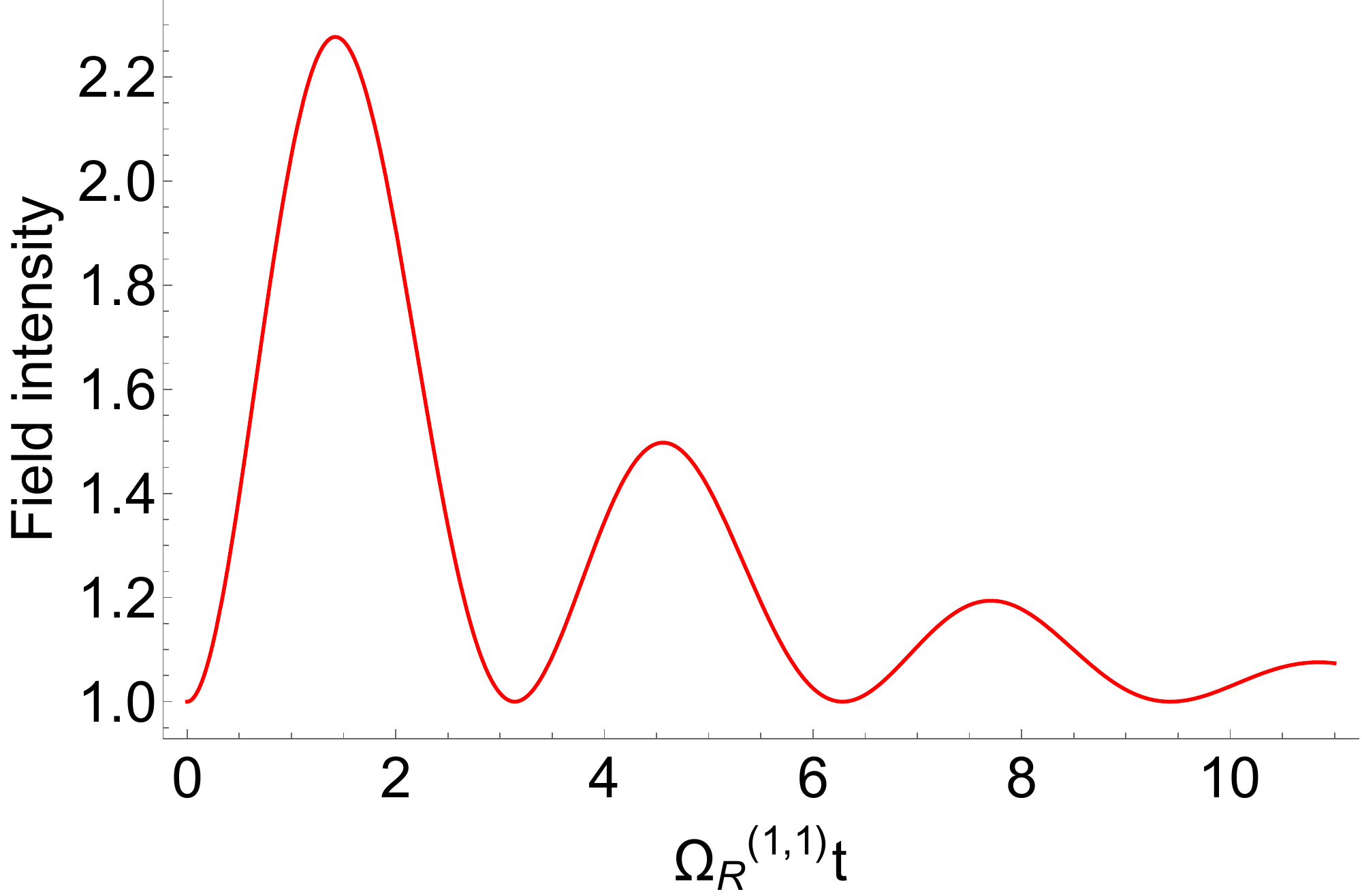}
\caption{ }
\label{fig4a}
\end{subfigure}

\begin{subfigure}[b]{0.5\textwidth}
\includegraphics[width=1\linewidth]{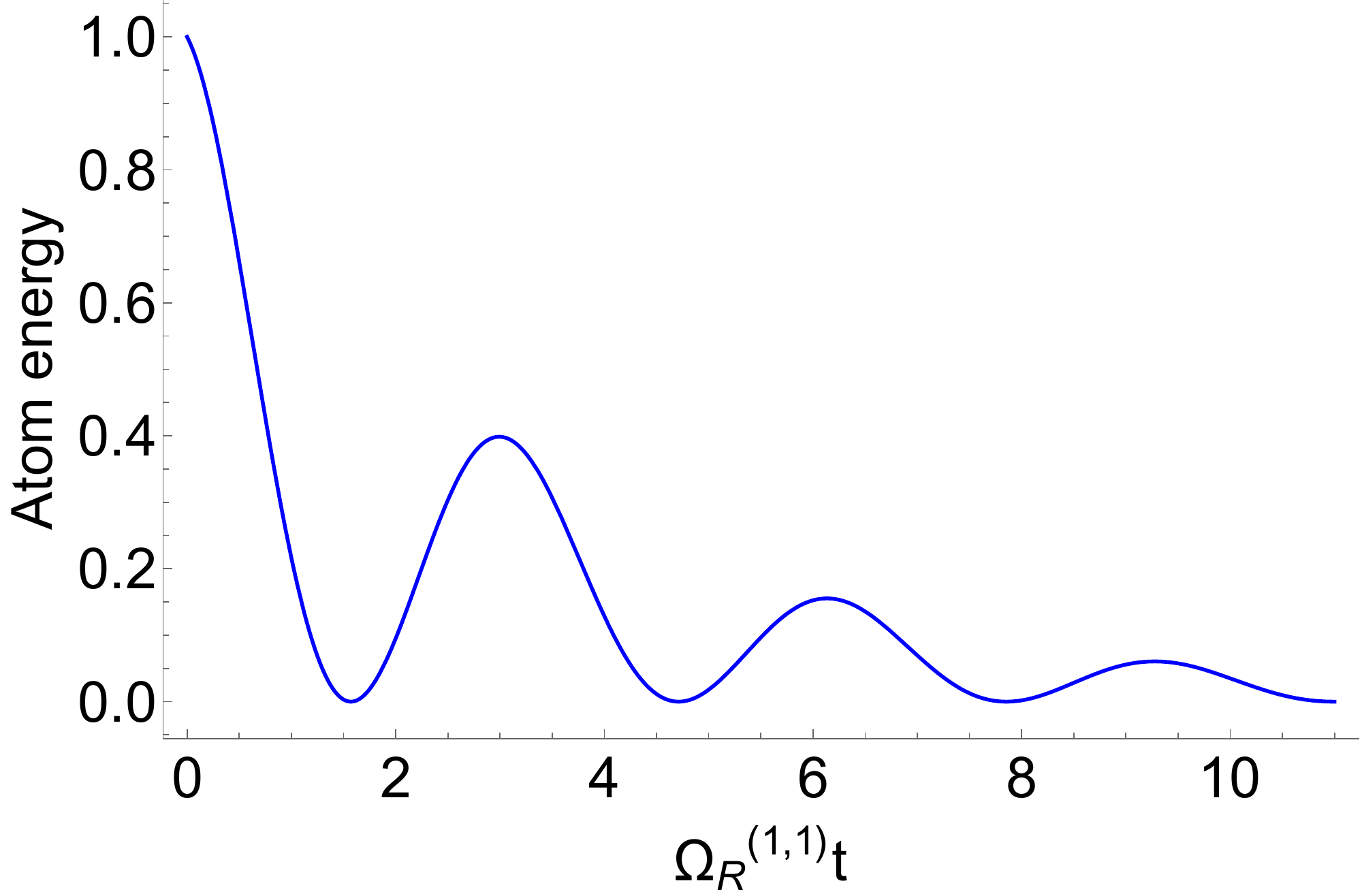}
\caption{  }
\label{fig4b}
\end{subfigure}
\caption{ (a) Normalized field intensity, $\left\langle \Psi \right\vert \mathbf{\hat{E}}^{2} \left\vert \Psi
\right\rangle / \left\vert \mathbf{E}\left( \mathbf{r}\right) \right\vert^{2}$, and  (b)  normalized atom energy $\left\langle \Psi \right\vert \hat{H}_{a}\left\vert \Psi \right\rangle /W$ as a function of time in units of the generalized Rabi frequency $\Omega _{R}^{\left( 1,1\right)}$. Here $ \gamma_{110}+\gamma _{001} = 0.3 \Omega _{R}^{\left( 1,1\right)}$.   }
\end{figure}


Fig. 4 illustrates the dynamics of observables in Eqs.~(\ref{modify av of e
square}) and~(\ref{modify av of ha}).

Note that the Weisskopf-Wigner theory would give the same expression~(\ref%
{modify av of ha}) for the atomic energy, but a wrong expression for the EM
field intensity:%
\begin{equation*}
\left\langle \Psi \right\vert \mathbf{\hat{E}}^{2}\left\vert \Psi
\right\rangle =\left\vert \mathbf{E}\left( \mathbf{r}\right) \right\vert ^{2}%
\left[ 2-\cos \left( 2\left\vert \widetilde{\Omega }_{R}^{\left( 1,1\right)
}\right\vert t\right) \right] e^{-\left( \gamma _{_{110}}+\gamma
_{001}\right) t},
\end{equation*}%
which does not approach the correct vacuum state.

\subsection{Emission spectra}

According to \cite{Scully1997}, the power spectrum of the emission is
\begin{align}
S(\mb{r},\nu) = \frac{1}{\pi} \mathrm{Re} \int_0^\infty d\tau G^{(1)}(\mb{r},\mb{r};\tau) e^{i\nu\tau} ,
\end{align} 
where $G^{(1)}(\mb{r},\mb{r};\tau)$ is the field autocorrelation function at the position $\mb{r}$ of the detector:
\begin{align}
G^{(1)}(\mb{r},\mb{r};\tau) = |\mb{E}(\mb{r})|^2 \int_0^\infty d t 
\overline { \langle \hat{c}_d^\dagger(t) \hat{c}_d(t+\tau) \rangle } .
\end{align}
where $\hat{c}_d(t),\hat{c}_d^\dagger(t)$ are annihilation and creation operators for the photons which interact with the detector, and the Heisenberg picture is used. We will assume that the coupling between the  photons and the detector is weak, so the photon detection does not affect the dynamics of the intracavity photons. According to \cite{madsen2013},  $\hat{c}_d(t) \propto \hat{c}(t)$ for a nanocavity, so we can calculate the $G^{(1)}(\mb{r},\mb{r};\tau)$ using operators for the cavity field $\hat{c}(t),\hat{c}^\dagger(t)$, up to a constant factor in the result. Note that the lower limit of the integral over $t$ is set to be $t=0$, which requires that no photons exist before $t=0$. 

In the Heisenberg-Langevin approach, an operator in the Heisenberg picture can be expressed through Schr\"{o}dinger's operators using the effective Hamiltonian $\hat{H}_{eff}$, which contains the anti-Hermitian part; see the Appendix. At the same time, the inhomogeneous term proportional to the noise sources should be added. Including these noise terms in the solution for the field operators when calculating the emission spectra is equivalent to taking into account the detection of thermal radiation which seeps into the cavity from outside and spontaneous emission resulting from thermal excitation of an atom. We assume that the reservoir temperature in energy units is much lower than $W$ and $\hbar \omega$, so that the contribution of these noise terms to the emission spectra can be neglected (although noise is still needed to preserve the norm). 

Then, the average correlator $\overline {\langle \hat{c}^\dagger(t) \hat{c}(t+\tau) \rangle}$ is expressed as
\begin{align}
&\phantom{{}={}} \overline{ \langle \hat{c}^\dagger(t) \hat{c}(t+\tau) \rangle }   \nonumber \\
&= \overline{ \langle \Psi(t=0) | e^{i\hat{H}_{eff}^\dagger t/\hbar} \hat{c}^\dagger e^{-i\hat{H}_{eff}t/\hbar} e^{i\hat{H}_{eff}^\dagger (t+\tau)/\hbar} \hat{c} e^{-i\hat{H}_{eff}(t+\tau)/\hbar} | \Psi(t=0) \rangle }  \nonumber \\
&= \overline{ \langle \Psi(t) | \hat{c}^\dagger e^{-i\hat{H}_{eff}t/\hbar} e^{i\hat{H}_{eff}^\dagger (t+\tau)/\hbar} \hat{c}  | \Psi(t+\tau) \rangle } ,
\end{align}
where $|\Psi(t)\rangle$ is the state vector of the system which we found in the previous subsection. It can be written as $|\Psi(t)\rangle = \sum_{n=0}^\infty C_{n}(t) |n\rangle |\Psi_n^{\alpha,e}(t)\rangle$, where $|\Psi_n^{\alpha,e}(t)\rangle$ is the part describing phonons and electrons. Therefore. 
\begin{align}
&\phantom{{}={}} \overline{ \langle \hat{c}^\dagger(t) \hat{c}(t+\tau) \rangle }   \nonumber \\
&= \overline{ \left( \sum_{n=0}^\infty C_{n}^\ast(t) \langle n | \langle \Psi_n^{\alpha,e}(t) | \right) \hat{c}^\dagger 
e^{-i\hat{H}_{eff}t/\hbar} e^{i\hat{H}_{eff}^\dagger (t+\tau)/\hbar} 
\hat{c} \left( \sum_{n=0}^\infty C_{n}(t+\tau) |n\rangle |\Psi_n^{\alpha,e}(t+\tau)\rangle \right) }     \nonumber \\
&= \overline{ \left( \sum_{n=0}^\infty \sqrt{n} C_{n}^\ast(t) \langle n-1 | \langle \Psi_n^{\alpha,e}(t) | \right)  
e^{-i\hat{H}_{eff}t/\hbar} e^{i\hat{H}_{eff}^\dagger (t+\tau)/\hbar}
\left( \sum_{n=0}^\infty \sqrt{n} C_{n}(t+\tau) |n-1\rangle |\Psi_n^{\alpha,e}(t+\tau)\rangle \right) } .
\end{align}

\begin{figure}[htb]
	\includegraphics[width=0.5\textwidth]{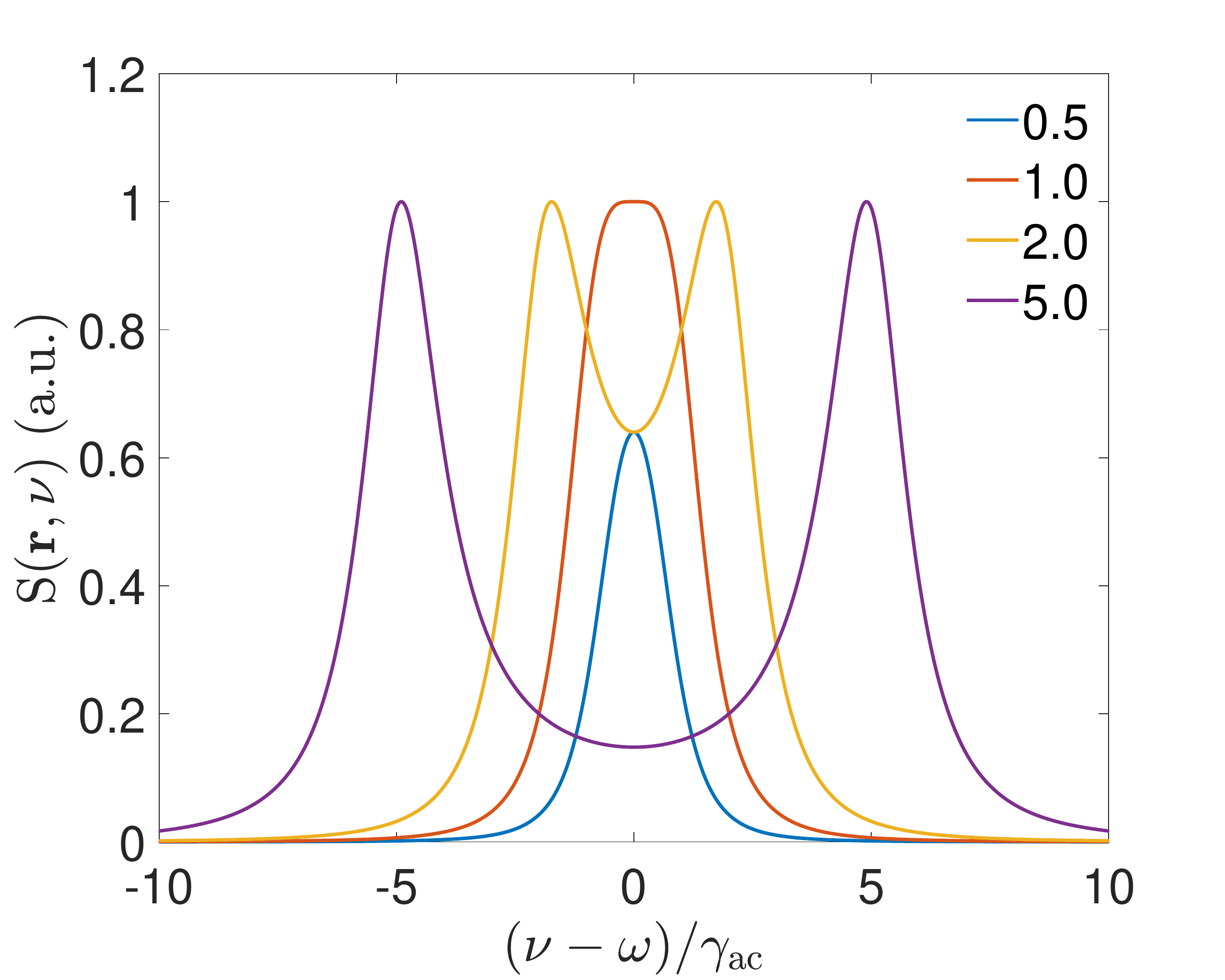}
	\caption{The emission spectra for $|\tilde{\Omega}_R^{(1,1)}| / \gamma_{\mathrm{ac}}$ equal to 0.5, 1, 2 and 5. All spectra are normalized by the same constant.  \label{Fig:power_spectrum}}
\end{figure}

Consider a simple example when the initial state is $| 0 \rangle | 0 \rangle | 1 \rangle$. Within the RWA  the system can only reach states $| 0 \rangle | 0 \rangle | 1 \rangle$, $| 1 \rangle | 1 \rangle | 0 \rangle$ and $| 0 \rangle | 0 \rangle | 0 \rangle$. After acting with $\hat{c}$ on a state of the system, a new state $| 1 \rangle | 0 \rangle | 0 \rangle$ can also appear, but it  cannot evolve into other states. So, in this case we have
\begin{align}
&\phantom{{}={}}\overline{ \langle \hat{c}^\dagger(t) \hat{c}(t+\tau) \rangle }  \nonumber \\
&= \overline{ \left( C_{1}^\ast(t) \langle 0 | \langle \Psi_1^{\alpha,e}(t) | \right)  
e^{-i\hat{H}_{eff}t/\hbar} e^{i\hat{H}_{eff}^\dagger (t+\tau)/\hbar}
\left( C_{1}(t+\tau) | 0 \rangle |\Psi_1^{\alpha,e}(t+\tau)\rangle \right) } \nonumber \\
&= \overline{ \left( C_{110}^\ast(t) \langle 1 | \langle 0 | \langle 0 | \right) 
e^{-i\hat{H}_{eff}t/\hbar} e^{i\hat{H}_{eff}^\dagger (t+\tau)/\hbar}
\left( C_{110}(t+\tau) | 1 \rangle | 0 \rangle | 0 \rangle \right) } \nonumber \\
&= C_{110}^\ast(t) C_{110}(t+\tau) \exp[ i \omega_{1,0} \tau  - \gamma_{100} (2t+\tau) ] ,
\end{align}
where we used Eqs.~(\ref{effectiv ham}) and (\ref{capital gamma}) and assumed that the noise for state $| 1 \rangle | 0 \rangle | 0 \rangle$ has zero correlator. Since 
\begin{align}
C_{110}(t) = i \sin\left( |\tilde{\Omega}_R^{(1,1)}| t \right) \exp[- i\omega_{1,1} t - \frac{\gamma_{110}+\gamma_{001}}{2} t  ] ,
\end{align}
we obtain
\begin{align}
\overline{ \langle \hat{c}^\dagger(t) \hat{c}(t+\tau) \rangle }  = 
\sin\left( |\tilde{\Omega}_R^{(1,1)}| t \right) 
\sin\left( |\tilde{\Omega}_R^{(1,1)}| (t+\tau) \right)
\exp[ -i\omega \tau]
\exp\left[- \gamma_{\mathrm{ac}} (2t+\tau) \right],
\end{align}
where we introduced the notation $\gamma_{\mathrm{ac}} \equiv \gamma_{100} + \frac{\gamma_{110}+\gamma_{001}}{2} $. 
Then the power spectrum is found to be
\begin{align}
S(\mb{r},\nu) 
\propto
\frac{1}{\pi} |\mb{E}(\mb{r})|^2 
\frac{|\tilde{\Omega}_R^{(1,1)}|^2} {4\gamma_{\mathrm{ac}} (|\tilde{\Omega}_R^{(1,1)}|^2+\gamma_{\mathrm{ac}}^2)}
\mathrm{Re} \left[ \frac{2\gamma_{\mathrm{ac}} - i (\nu-\omega)} { \left[ \gamma_{\mathrm{ac}}-i(\nu-\omega) \right]^2+|\tilde{\Omega}_R^{(1,1)}|^2 } \right] .
\end{align}
The normalized power spectra are shown in Fig.~\ref{Fig:power_spectrum} for various values of $|\tilde{\Omega}_R^{(1,1)}| / \gamma_{\mathrm{ac}}$. For $|\tilde{\Omega}_R^{(1,1)}| < \gamma_{\mathrm{ac}}$ the spectrum has a single maximum at zero detuning $\nu = \omega$. For $|\tilde{\Omega}_R^{(1,1)}| > \gamma_{\mathrm{ac}}$ the spectra are split and their maxima (same value for all spectra) are reached at detunings given by $(\nu-\omega)^2 = |\tilde{\Omega}_R^{(1,1)}|^2 - \gamma_{\mathrm{ac}}^2$.  Therefore, to reach the strong coupling regime the Rabi frequency $|\tilde{\Omega}_R^{(1,1)}|$ has to exceed the combination of the decoherence rates denoted by  $\gamma_{\mathrm{ac}}$. 


\subsection{Relaxation rates}

Finally, we give explicit expressions for the relaxation constants $\gamma
_{\alpha n0}$ and $\gamma _{\alpha n1}$. They were derived in Appendix
using the Lindblad master equation approach and assuming statistical
independence of ``partial'' dissipative
reservoirs for the atomic, EM, and phonon subsystems. The result is%
\begin{equation}
\gamma _{\alpha n0}=\frac{\gamma }{2}N_{1}^{T_{a}}+\frac{\mu _{\omega }}{2}%
\left[ \overline{n}_{\omega }^{T_{em}}\left( n+1\right) +\left( \overline{n}%
_{\omega }^{T_{em}}+1\right) n\right] +\frac{\mu _{\Omega }}{2}\left[
\overline{n}_{\Omega }^{T_{p}}\left( \alpha +1\right) +\left( \overline{n}%
_{\Omega }^{T_{p}}+1\right) \alpha \right] ,  \label{gamma0}
\end{equation}

\begin{equation}
\gamma _{\alpha n1}=\frac{\gamma }{2}N_{0}^{T_{a}}+\frac{\mu _{\omega }}{2}%
\left[ \overline{n}_{\omega }^{T_{em}}\left( n+1\right) +\left( \overline{n}%
_{\omega }^{T_{em}}+1\right) n\right] +\frac{\mu _{\Omega }}{2}\left[
\overline{n}_{\Omega }^{T_{p}}\left( \alpha +1\right) +\left( \overline{n}%
_{\Omega }^{T_{p}}+1\right) \alpha \right] ,  \label{gamma1}
\end{equation}%
where $\gamma $, $\mu _{\omega }$ and $\mu _{\Omega }$ are partial
relaxation rates of the atomic, photon, and phonon subsystems respectively; $%
N_{0}^{T_{a}}=\frac{1}{1+e^{-\frac{W}{T_{a}}}}$, $N_{1}^{T_{a}}=\frac{^{e^{-%
\frac{W}{T_{a}}}}}{1+e^{-\frac{W}{T_{a}}}}$, $\overline{n}_{\omega
}^{T_{em}}=$ $\frac{1}{e^{\frac{\hbar \omega }{T_{em}}}-1}$, $\overline{n}%
_{\Omega }^{T_{p}}=\frac{1}{e^{\frac{\hbar \Omega }{Tp}}-1}$ are their
occupation numbers at thermal equilibrium; $T_{a,em,p}$ are temperatures
of partial atom, photon, and phonon reservoirs in energy units. As a reminder, the atom energy is equal to $0$ in
state $\left\vert 0\right\rangle $ and $W$ in state $\left\vert
1\right\rangle $ .

If all reservoirs are at zero temperature, we obtain%
\begin{equation}
\gamma _{\alpha n0}=\frac{\mu _{\omega }}{2}n+\frac{\mu _{\Omega }}{2}\alpha
, \; \gamma _{\alpha n1}=\frac{\gamma }{2}+\frac{\mu _{\omega }}{2}n+\frac{\mu
_{\Omega }}{2}\alpha .  \label{gamma at 0 tem}
\end{equation}%
Eq.~(\ref{gamma at 0 tem}) shows that$\ \gamma _{000}=0$, validating our
choice earlier in this section. We also obtain physically intuitive
expressions for $\gamma _{110}$ and $\gamma _{001}$ : $\gamma _{110}=\frac{%
\mu _{\omega }}{2}+\frac{\mu _{\Omega }}{2}$, $\gamma _{001}=\frac{\gamma }{2%
}$.


\section{Classical acoustic pumping}

In this case the RWA Hamiltonian is given by Eq.~(\ref{sch}). It depends
only on quantum operators $\hat{\sigma},\hat{\sigma}^{\dagger }$ and $\hat{c}%
,\hat{c}^{\dagger }$; therefore the state vector has to be expanded over the
basis states $\left\vert n\right\rangle \left\vert 0\right\rangle $ and $%
\left\vert n\right\rangle \left\vert 1\right\rangle $:%
\begin{equation}
\Psi =\sum_{n=0}^{\infty }\left( C_{n0}\left\vert n\right\rangle \left\vert
0\right\rangle +C_{n1}\left\vert n\right\rangle \left\vert 1\right\rangle
\right) .  \label{scpsi}
\end{equation}%
Substituting Eq.~(\ref{scpsi}) in the Schr\"{o}dinger equation with the
Hamiltonian~(\ref{sch}), we again get separation into a block of two
equations,%
\begin{equation}
\dot{C}_{n0}=-i\omega _{n}C_{n0}+i\frac{\mathfrak{R}^{\ast }}{\hbar }e^{i\Omega
t}C_{\left( n-1\right) 1}\sqrt{n},  \label{cn0 dot}
\end{equation}
\begin{equation}
\dot{C}_{\left( n-1\right) 1}=-i\left( \omega _{n-1}+\frac{W}{\hbar }\right)
C_{\left( n-1\right) 1}+i\frac{\mathfrak{R}}{\hbar }e^{-i\Omega t}C_{n0}\sqrt{n},
\label{cn-11 dot}
\end{equation}%
and a separate equation for the amplitude of the ground state  $%
\left\vert 0\right\rangle \left\vert 0\right\rangle $ of the system:
\begin{equation} 
\dot{C}_{00}=-i\omega _{0}C_{00}, 
 \label{c00 dot}
\end{equation}%
where $\omega _{n}=\omega \left( n+\frac{1}{2}\right) $. After making the substitution 
 $C_{\left( n-1\right) 1}=G_{\left( n-1\right) 1}e^{-i\Omega t}$
, Eqs.~(\ref{cn0 dot}),(\ref{cn-11 dot}) give the equations similar in form to
Eqs.~(\ref{coeff eq for psi}):%
\begin{equation}
\frac{d}{dt}\left(
\begin{array}{c}
C_{n0} \\
G_{\left( n-1\right) 1}%
\end{array}%
\right) +\left(
\begin{array}{cc}
i\omega _{n} & -i\Omega _{R}^{\left( n\right) \ast } \\
-i\Omega _{R}^{\left( n\right) } & i\omega _{n}-i\Delta%
\end{array}%
\right) \left(
\begin{array}{c}
C_{n0} \\
G_{\left( n-1\right) 1}%
\end{array}%
\right) =0,  \label{sc coeff eq for psi}
\end{equation}%
where%
\begin{equation*}
\Omega _{R}^{\left( n\right) }=\frac{\mathfrak{R}}{\hbar }\sqrt{n}, \; \Delta =\Omega
+\omega -\frac{W}{\hbar }, \; \omega _{n}-\Delta =\omega _{n-1}+\frac{W}{\hbar }.
\end{equation*}%
Eqs.~(\ref{c00 dot}),(\ref{sc coeff eq for psi}) are different from Eqs.~(%
\ref{coeff eq for psi}),(\ref{eq for les}) only in one aspect: they don't
contain the index of the quantum state of the phonon field, whereas the
Rabi frequency depends on the amplitude of classical acoustic oscillations $%
\mathbf{Q}\left( \mathbf{r}_{a}\right)$, see Sec.~IIE. Obviously, the solution to Eqs.~(%
\ref{c00 dot}),(\ref{sc coeff eq for psi}) will have the same form and the
expressions (\ref{av of e square}),~(\ref{av of ha}) for the observables
will remain the same, after dropping the index of the quantum phonon state
and redefining the Rabi frequency.

Dissipation due to coupling to a reservoir can be included using the
stochastic equation of evolution of the state vector, see the Appendix. The corresponding equations are
again similar to those for a fully quantum problem given by Eqs.~(\ref{modify coeff eq
for psi}),(\ref{modify eq for les}):
\begin{equation}
\dot{C}_{00}+i\left( \omega _{0}+\gamma _{00}\right) C_{00}=-\frac{i}{\hbar }%
R_{00},  \label{modify c00 dot}
\end{equation}
\begin{equation}
\frac{d}{dt}\left(
\begin{array}{c}
C_{n0} \\
C_{\left( n-1\right) 1}%
\end{array}%
\right) +\left(
\begin{array}{cc}
i\omega _{n}+\gamma _{n0} & -i\Omega _{R}^{\left( n\right) \ast } \\
-i\Omega _{R}^{\left( n\right) } & i\omega _{n}-i\Delta +\gamma _{\left(
n-1\right) 1}%
\end{array}%
\right) \left(
\begin{array}{c}
C_{n0} \\
C_{\left( n-1\right) 1}%
\end{array}%
\right) =-\frac{i}{\hbar }\left(
\begin{array}{c}
R_{n0} \\
R_{\left( n-1\right) 1}%
\end{array}%
\right). 
 \label{modify sc coeff eq for psi}
\end{equation}%
Since the acoustic field is now a given external pumping, the relaxation
constants should not depend on the parameters of a phonon reservoir. They
can be obtained after obvious simplification of Eqs.~(\ref{gamma0}),(\ref%
{gamma1}):

\begin{equation}
\gamma _{n0}=\frac{\gamma }{2}N_{1}^{T_{a}}+\frac{\mu _{\omega }}{2}\left[
\overline{n}_{\omega }^{T_{em}}\left( n+1\right) +\left( \overline{n}%
_{\omega }^{T_{em}}+1\right) n\right] ,  \label{sc gamma0}
\end{equation}

\begin{equation}
\gamma _{n1}=\frac{\gamma }{2}N_{0}^{T_{a}}+\frac{\mu _{\omega }}{2}\left[
\overline{n}_{\omega }^{T_{em}}\left( n+1\right) +\left( \overline{n}%
_{\omega }^{T_{em}}+1\right) n\right] ,  \label{sc gamma1}
\end{equation}%
All expressions for the state vector and observables can be obtained from
the corresponding expressions in Sec.~IV after dropping the index $\alpha $
of the quantum state of the phonon field and redefining the frequency of
Rabi oscillations.

\section{Separation and interplay of the parametric and one-photon resonance}

For an electron system coupled to a EM cavity mode and dressed by a phonon
field, the phonon frequency $\Omega $ can be much lower than the optical
frequency. In this case the overlap of the parametric (three-wave) resonance
$\omega \pm \Omega \approx \frac{W}{\hbar }$ and the one-photon (two-wave)
resonance $\omega \approx \frac{W}{\hbar }$ can be an issue.

First of all, it is clear that the resonances can be separated only
if the value of $\Omega $ exceeds the sum of the spectral widths of the EM
cavity mode and the electron transition.

Second, the separation criterion imposes certain restrictions on the
Rabi frequencies of the two resonances. To derive these restrictions, we
neglect dissipation and retain in the Hamiltonian~(\ref{toh}) both the RWA
terms near the parametric resonance $\omega \pm \Omega \approx \frac{W}{%
\hbar }$, and the terms near a one-photon resonance $\omega \approx \frac{W}{%
\hbar }$. Since the result will be almost the same whether the phonon field
is quantized or classical, we will consider the classical phonon field to
keep the expressions a bit shorter. The resulting Hamiltonian is%
\begin{equation}
\hat{H}=\hbar \omega \left( \hat{c}^{\dagger }\hat{c}+\frac{1}{2}\right) +W%
\hat{\sigma}^{\dagger }\hat{\sigma}-\left( \chi + \mathfrak{R} e^{-i\Omega t}\right) \hat{%
\sigma}^{\dagger }\hat{c}-\left( \chi ^{\ast }+ \mathfrak{R}^{\ast }e^{i\Omega t}\right)
\hat{\sigma}\hat{c}^{\dagger },  \label{rh}
\end{equation}%
where\ $\chi =\left( \mathbf{d}\cdot \mathbf{E}\right) _{\mathbf{r}=\mathbf{r%
}_{a}},\mathfrak{R} =\left[ \mathbf{d}\left( \mathbf{Q\cdot \nabla }\right) \mathbf{E}%
\right] _{\mathbf{r}=\mathbf{r}_{a}}$; $\mathbf{Q}$ is now a
complex-valued amplitude of classical phonon oscillations. The value of $%
\Omega $ in Eq.~(\ref{rh}) can be both positive and negative, corresponding
to the choice of an upper or lower sign in the parametric resonance
condition $\omega \pm \Omega \approx \frac{W}{\hbar }$. The change of sign
in $\Omega $ corresponds to replacing $\mathbf{Q}$ with $\mathbf{Q}^{\ast }$
in the expression for $\mathfrak{R}$.

The state vector should be sought in the form of Eq.~(\ref{scpsi}). After
substituting it into the Schr\"{o}dinger equation we obtain coupled
equations for the amplitudes of basis states $\left\vert n\right\rangle
\left\vert 0\right\rangle $, $\left\vert n-1 \right\rangle \left\vert
1\right\rangle $:%
\begin{align}
	& \dot{C}_{n0} + i\omega_{n}C_{n0} - \frac{i}{\hbar }\left( \chi^{\ast } + \mathfrak{R}^{\ast}e^{i\Omega t}\right) \sqrt{n} C_{\left( n-1\right) 1} = 0,  \label{eq for cn0}  \\
	& \dot{C}_{\left( n-1\right) 1} + i\left( \omega_{n-1} + \frac{W}{\hbar} \right)	C_{\left( n-1\right) 1} - \frac{i}{\hbar} \left( \chi + \mathfrak{R} e^{-i\Omega t}\right)
	\sqrt{n} C_{n0} = 0, 
	 \label{eq for cn-11}
\end{align}
and 
\begin{equation}
\dot{C}_{00}+i\omega _{0}C_{00}=0,  
\label{eq for c00}
\end{equation}
where $\omega _{n}=\omega \left( n+\frac{1}{2}\right) $. To compare these
equations with Eqs.~(\ref{cn0 dot}) and~(\ref{cn-11 dot}), it is convenient
to assume that the system is exactly at one of the resonances and study the
behavior of the solution with increasing the detuning from another
resonance. For example, we assume an exact parametric resonance $\omega
+\Omega =\frac{W}{\hbar }$. In this case the detuning from the two-wave
resonance is $\frac{W}{\hbar }-\omega =\Omega $. After the substitution $C_{n0}=G_{n0}e^{-i\omega _{n}t}$ and $C_{\left( n-1\right)
1}=G_{\left( n-1\right) 1}e^{-i\left( \omega _{n-1}+\frac{W}{\hbar }\right)
t},$ we obtain from Eqs.~(\ref{eq for cn0}) and~(\ref{eq for cn-11}) that 
\begin{align}
& \dot{G}_{n0} - \frac{i}{\hbar }\left( \chi^{\ast } + \mathfrak{R}^{\ast}e^{i\Omega t}\right) \sqrt{n} G_{\left( n-1\right) 1} e^{-i \left( \frac{W}{\hbar} - \omega \right) t} = 0,   \label{gn0 dot}  \\
& \dot{G}_{\left( n-1\right) 1}  - \frac{i}{\hbar} \left( \chi + \mathfrak{R} e^{-i\Omega t}\right)
\sqrt{n} G_{n0} e^{i \left( \frac{W}{\hbar} - \omega \right) t} = 0 .    \label{gn-11 dot}
\end{align}

If we neglect at first the perturbation of the system in the vicinity of the
two-wave resonance, the solution to Eqs.~(\ref{gn0 dot}) and~(\ref{gn-11 dot}) at $\chi =0$ is%
\begin{equation}
\left(
\begin{array}{c}
G_{n0} \\
G_{\left( n-1\right) 1}%
\end{array}%
\right) =Ae^{i\Omega _{R}^{\left( 3\right) }t}\left(
\begin{array}{c}
1 \\
1%
\end{array}%
\right) + Be^{-i\Omega _{R}^{\left( 3\right) }t}\left(
\begin{array}{c}
1 \\
-1%
\end{array}%
\right) ,  \label{sol to gn0 and gn-11}
\end{equation}%
where $\Omega _{R}^{\left( 3\right) }=\frac{1}{\hbar }\mathfrak{R} \sqrt{n}$ is the Rabi
frequency of the parametric resonance, $A$, and $B$ are arbitrary constants.
The state described by Eq.~(\ref{sol to gn0 and gn-11}) is obviously
entangled.

To write the formal solution to Eqs.~(\ref{gn0 dot}) and~(\ref{gn-11 dot}),
we make another substitution of variables: $G_{n0}\pm G_{\left( n-1\right)
1}=G_{\pm }$ . The result is%
\begin{equation*}
\dot{G}_{\pm }\mp i\Omega _{R}^{\left( 3\right) }G_{\pm }=i\Omega
_{R}^{\left( 2\right) \ast }e^{-i\Omega t}G_{n0}\pm i\Omega _{R}^{\left(
2\right) }e^{i\Omega t}G_{\left( n-1\right) 1},
\end{equation*}%
where $\Omega _{R}^{\left( 2\right) }=\frac{1}{\hbar }\chi \sqrt{n}$ is the
Rabi frequency corresponding to the one-photon (two-wave) resonance. The
solution to the last equation is
\begin{equation}
G_{\pm }= 2(A,B) e^{\pm i\Omega _{R}^{\left( 3\right) }t}+ie^{\pm
i\Omega _{R}^{\left( 3\right) }t}\int_{0}^{t}e^{\mp i\Omega _{R}^{\left(
3\right) }\tau }\left[ \Omega _{R}^{\left( 2\right) \ast }e^{-i\Omega \tau
}G_{n0}\left( \tau \right) \pm \Omega _{R}^{\left( 2\right) }e^{i\Omega \tau
}G_{\left( n-1\right) 1}\left( \tau \right) \right] d\tau .  \label{g+-}
\end{equation}%
Considering the terms proportional to $\Omega _{R}^{\left( 2\right) }$as
perturbation, we seek the solution as%
\begin{equation*}
\left(
\begin{array}{c}
G_{n0} \\
G_{\left( n-1\right) 1}%
\end{array}%
\right) =Ae^{i\Omega _{R}^{\left( 3\right) }t}\left(
\begin{array}{c}
1 \\
1%
\end{array}%
\right) + Be^{-i\Omega _{R}^{\left( 3\right) }t}\left(
\begin{array}{c}
1 \\
-1%
\end{array}%
\right) +\left(
\begin{array}{c}
\delta G_{n0} \\
\delta G_{\left( n-1\right) 1}%
\end{array}%
\right) .
\end{equation*}%
To estimate the magnitude of the perturbation, we substitute Eq.~(\ref{sol
to gn0 and gn-11}) into Eq.~(\ref{g+-}). After some algebra we obtain that 
under the condition $\Omega _{R}^{\left( 3\right)
}\ll \Omega $ the magnitude of the perturbation is%
\begin{equation*}
\delta G_{n0,\left( n-1\right) 1}\sim \left\vert \frac{\Omega _{R}^{\left(
2\right) }}{\Omega }\right\vert G_{n0,\left( n-1\right) 1},
\end{equation*}%
whereas if $\Omega _{R}^{\left( 3\right) }\sim \Omega $ the magnitude of the
perturbation is%
\begin{equation*}
\delta G_{n0,\left( n-1\right) 1}\sim \left\vert \frac{\Omega _{R}^{\left(
2\right) }}{\Omega _{R}^{\left( 3\right) }}\right\vert G_{n0,\left(
n-1\right) 1}.
\end{equation*}

To summarize this part, if both Rabi frequencies $\Omega _{R}^{\left(
3\right) },$ $\Omega _{R}^{\left( 2\right) }\ll \Omega $, the two resonances
can be treated independently for any relationship between the magnitudes of $%
\Omega _{R}^{\left( 3\right) }$ and $\Omega _{R}^{\left( 2\right) }$. If the
above inequality is violated, one can neglect one of the resonances only if
its associated Rabi frequency is much lower than the Rabi frequency of
another resonance. These restrictions are obvious from qualitative physical
reasoning: either the magnitudes of the Rabi splittings are much smaller
than the distance between resonances, or one of the splittings is much
weaker than another one.

When the effect of the neighboring resonance is non-negligible, it can still
be taken into account in the solution. Indeed, consider the solution to
Eqs.~(\ref{eq for cn0}) and~(\ref{eq for cn-11}), taking into account only
the two-wave resonance, i.e. taking $\mathfrak{R} =0$. After obvious substitutions, we
arrive at%
\begin{eqnarray}
\left(
\begin{array}{c}
C_{n0} \\
C_{\left( n-1\right) 1}%
\end{array}%
\right) &=&Ae^{-i\left( \omega _{n}-\frac{\Omega }{2}+\sqrt{\frac{\Omega ^{2}%
}{4}+\left\vert \Omega _{R}^{\left( 2\right) }\right\vert ^{2}}\right)
t}\times \left(
\begin{array}{c}
1 \\
\frac{-\frac{\Omega }{2}+\sqrt{\frac{\Omega ^{2}}{4}+\left\vert \Omega
_{R}^{\left( 2\right) }\right\vert ^{2}}}{\Omega _{R}^{\left( 2\right) }}%
\end{array}%
\right)  \notag \\
&&+Be^{-i\left( \omega _{n-1}+\frac{W}{\hbar }+\frac{\Omega }{2}-\sqrt{\frac{%
\Omega ^{2}}{4}+\left\vert \Omega _{R}^{\left( 2\right) }\right\vert ^{2}}%
\right) t}\times \left(
\begin{array}{c}
\frac{\Omega _{R}^{\left( 2\right) }}{-\frac{\Omega }{2}-\sqrt{\frac{\Omega
^{2}}{4}+\left\vert \Omega _{R}^{\left( 2\right) }\right\vert ^{2}}} \\
1%
\end{array}%
\right) ;  \label{cn0 and cn-11}
\end{eqnarray}%
In the limit $\Omega \gg \Omega _{R}^{\left( 2\right) }$ we obtain%
\begin{equation}
\left(
\begin{array}{c}
C_{n0} \\
C_{\left( n-1\right) 1}%
\end{array}%
\right) \approx Ae^{-i\left( \omega _{n}+\frac{\left\vert \Omega
_{R}^{\left( 2\right) }\right\vert ^{2}}{\Omega }\right) t}\left(
\begin{array}{c}
1 \\
\frac{\left\vert \Omega _{R}^{\left( 2\right) }\right\vert }{\Omega }%
\end{array}%
\right) +Be^{-i\left( \omega _{n-1}+\frac{W}{\hbar }-\frac{\left\vert \Omega
_{R}^{\left( 2\right) }\right\vert ^{2}}{\Omega }\right) t}\left(
\begin{array}{c}
-\frac{\left\vert \Omega _{R}^{\left( 2\right) }\right\vert }{\Omega } \\
1%
\end{array}%
\right)  \label{lim cn0 and cn-11}
\end{equation}%
It is clear from Eq.~(\ref{lim cn0 and cn-11}) that the entanglement of
states described by $C_{n0}$ and $C_{\left( n-1\right) 1}$ is determined by
a small parameter $\frac{\left\vert \Omega _{R}^{\left( 2\right)
}\right\vert }{\Omega }$ , whereas at exact resonance the entanglement is
always stronger; see Eq.~(\ref{sol to gn0 and gn-11}). Therefore, when $\Omega
_{R}^{\left( 2\right) }\ll \Omega $, we can neglect the contribution of the
two-photon resonance to the entanglement of states $\left\vert
n\right\rangle \left\vert 0\right\rangle $ and $\left\vert n-1\right\rangle
\left\vert 1\right\rangle $. However, it follows from Eq.~(\ref{cn0 and
cn-11}) that the two-wave resonance shifts the eigenfrequencies of the
system. Qualitatively, these shifts can be included by putting $\chi =0$ in
Eqs.~(\ref{eq for cn0}) and~(\ref{eq for cn-11}) but replacing the
eigenfrequencies $\omega _{n}$ and $\omega _{n-1}$ according to Eq.~(\ref%
{lim cn0 and cn-11}):%
\begin{equation}
\omega _{n}\Longrightarrow \omega _{n}+\frac{\left\vert \Omega _{R}^{\left(
2\right) }\right\vert ^{2}}{\Omega }, \; \omega _{n-1}+\frac{W}{\hbar } 
\Longrightarrow \omega _{n-1}+\frac{W}{\hbar }-\frac{\left\vert \Omega
_{R}^{\left( 2\right) }\right\vert ^{2}}{\Omega }.  \label{shift omega}
\end{equation}%
If $\Omega _{R}^{\left( 3\right) }\ll \frac{\left\vert \Omega _{R}^{\left(
2\right) }\right\vert ^{2}}{\Omega }$ these shifts can be significant in
order to interpret the spectra near the three-wave parametric resonance.

The same reasoning can be carried out to analyze the effect of a detuned
three-wave resonance on the solution near the two-wave resonance.

These results can be verified by an exact numerical solution of Eqs.~(\ref{eq for cn0}) and~(\ref{eq for cn-11}) for given initial conditions.   After that, we can obtain the spectra of $C_{n0}$ and $C_{(n-1)1}$. Since they are oscillating functions, their spectra form discrete lines at frequencies which we denote as 
$\omega_{\rm osc}$. 

\begin{figure}[htb]
	\centering
	\begin{subfigure}{0.45\textwidth}
		\centering
		\includegraphics[width=\linewidth]{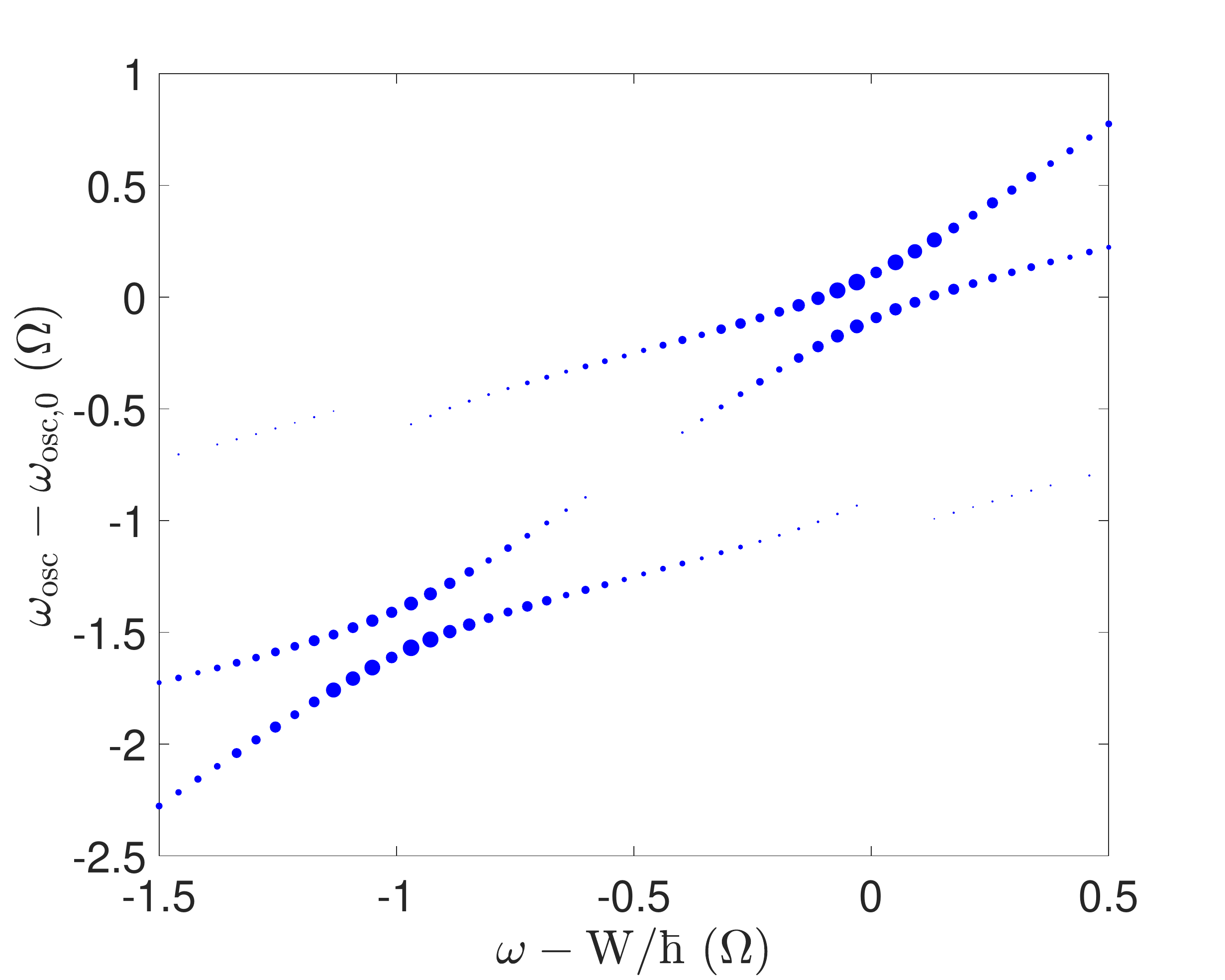}
	\end{subfigure}
	\begin{subfigure}{0.45\textwidth}
		\includegraphics[width=\linewidth]{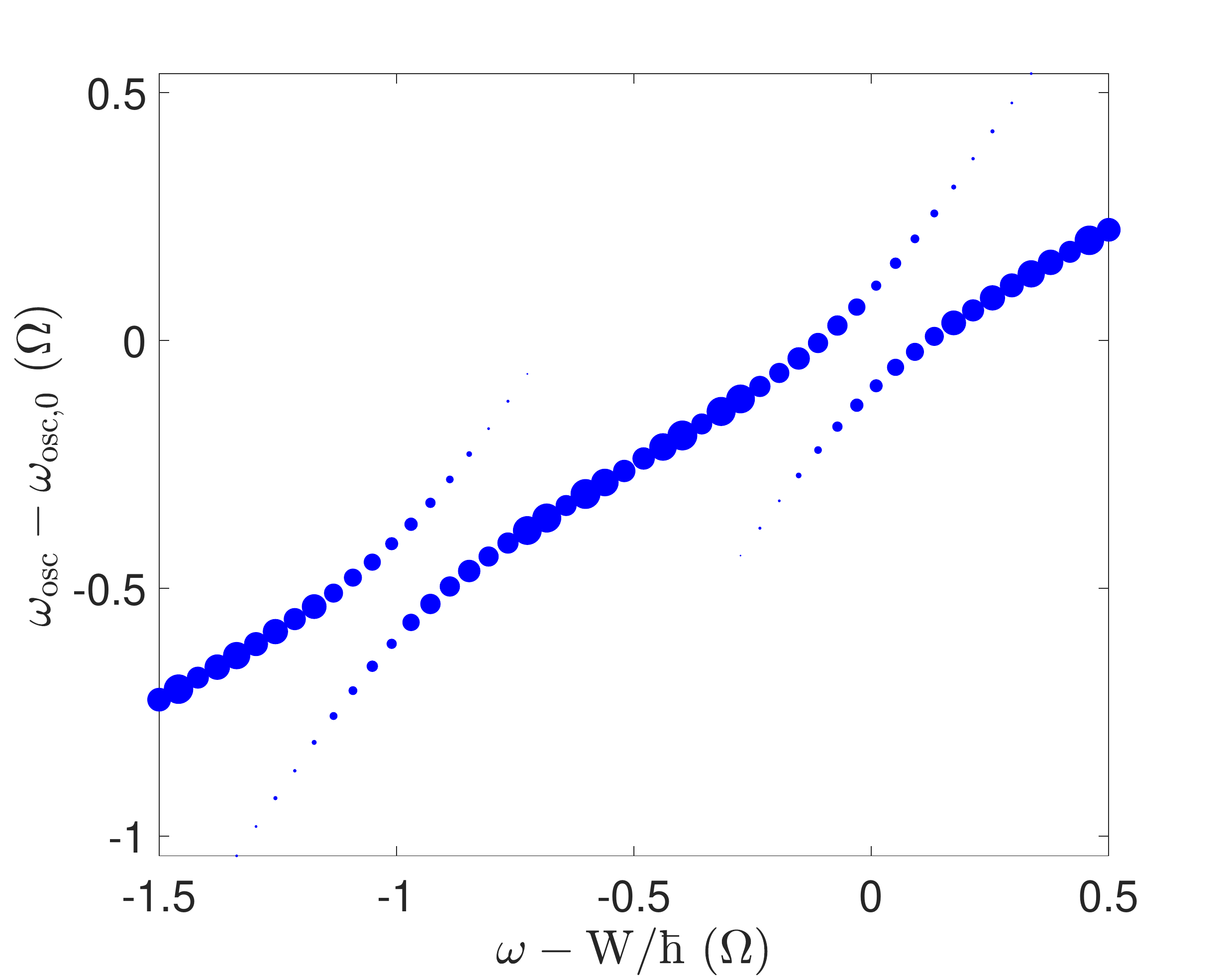}
	\end{subfigure}
	\caption{ The frequencies $\omega_{\rm osc}$ of the spectral lines for $C_{n0}$ (left panel), and $C_{(n-1)1}$ (right panel), with $n=1$,  as functions of the photon frequency $\omega$. The photon frequencies are shifted by $W/\hbar$, and the positions of spectral lines $\omega_{\rm osc}$ are shifted by $\omega_{osc,0} = \omega_1|_{\omega=W/\hbar}$. The area of a marker is proportional to the amplitude squared of the spectral line. Both axes are in units of $\Omega$. The parameters are $|\Omega_R^{(2)}| = |\Omega_R^{(3)}| = 0.1\Omega$, and the initial condition is $C_{n0}(0)=0$ and $C_{(n-1)1}(0)=1$.} 
	\label{Fig:resonance_separation}
\end{figure}

As an example, we select the case of $n=1$, set $|\Omega_R^{(2)}| = |\Omega_R^{(3)}| = 0.1\Omega$, and choose the initial condition as $C_{n0}(0)=0$ and $C_{(n-1)1}(0)=1$. The frequencies $\omega_{\rm osc}$ of the spectral lines for $C_{n0}$ and $C_{(n-1)1}$ are shown in Fig.~\ref{Fig:resonance_separation}. Their values are shifted by $\omega_{osc,0} = \omega_1|_{\omega=W/\hbar}$. The area of the dot for each spectral line is proportional to the square of its amplitude. If a marker is not visible, it means the corresponding line is very weak and can be neglected. The anticrossing can be seen at both the one-photon resonance and parametric resonance.

\begin{figure}[htb]
	\centering
	\begin{subfigure}{0.45\textwidth}
		\centering
		\includegraphics[width=\linewidth]{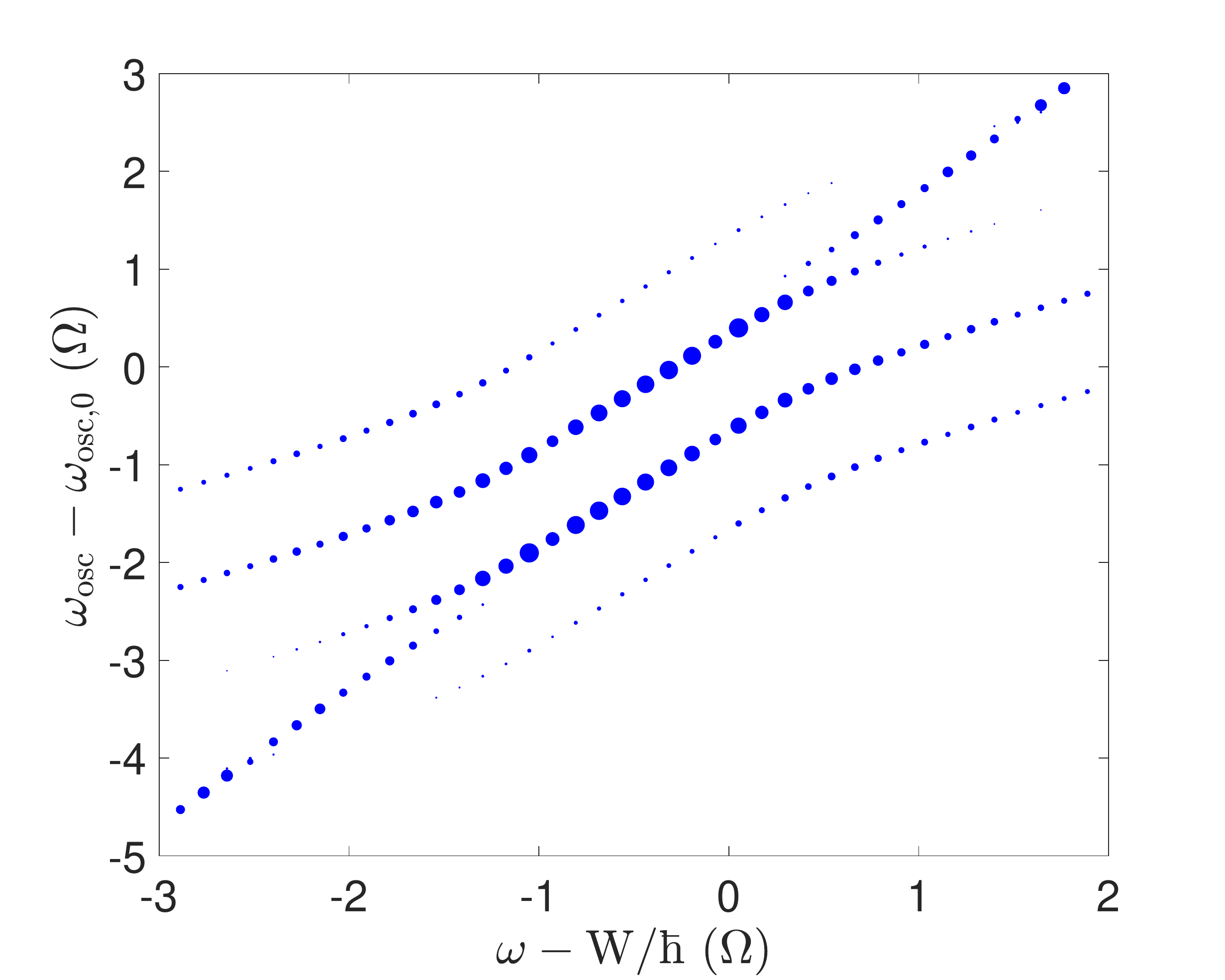}
	\end{subfigure}
	\begin{subfigure}{0.45\textwidth}
		\includegraphics[width=\linewidth]{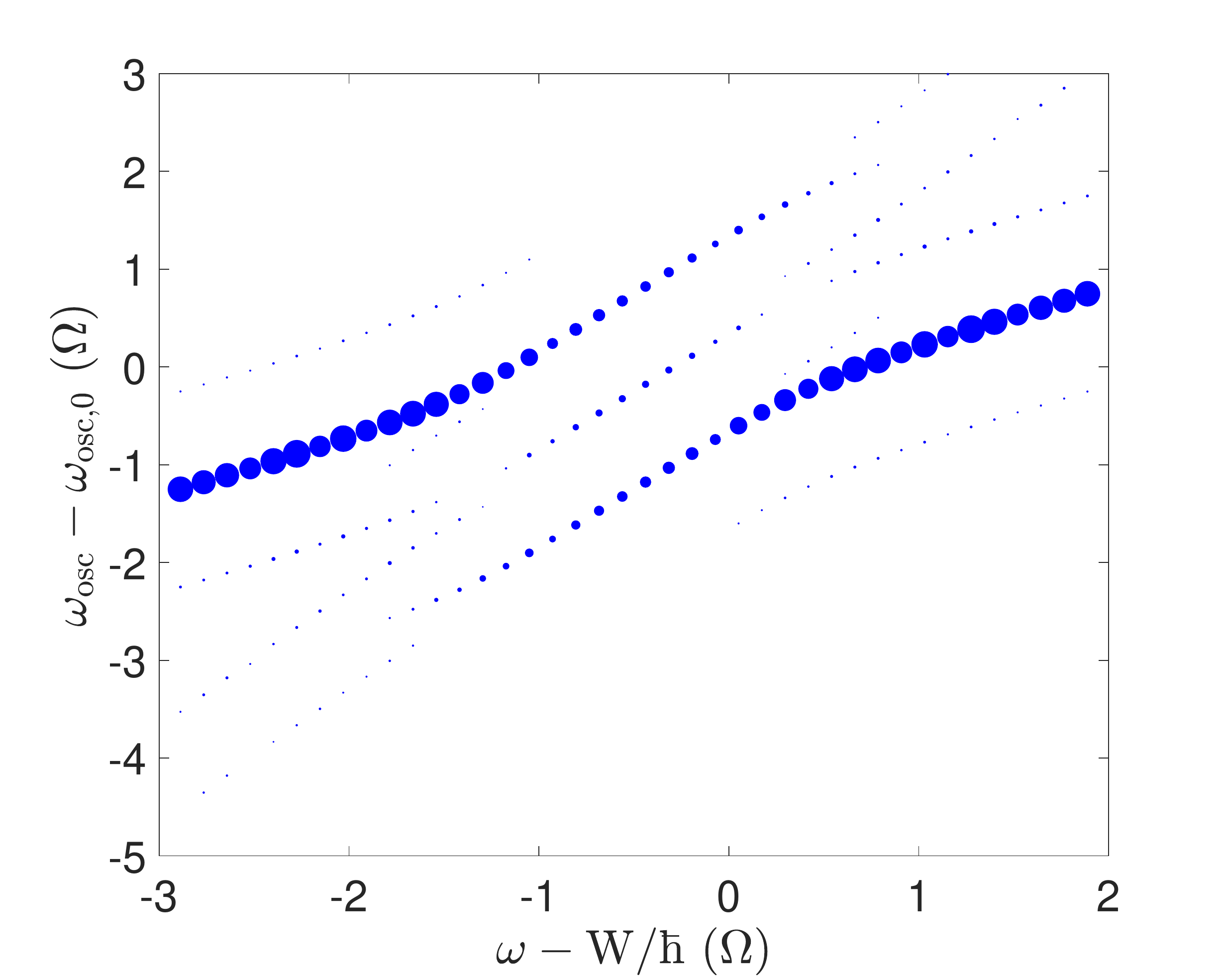}
	\end{subfigure}
	\caption{ The frequencies $\omega_{\rm osc}$ of the spectral lines for $C_{n0}$ (left panel), and $C_{(n-1)1}$ (right panel),  with $n=1$. The notations are the same as in Fig.~\ref{Fig:resonance_separation}. The parameters are $|\Omega_R^{(2)}| = |\Omega_R^{(3)}| = 0.5\Omega$, and the initial condition is $C_{n0}(0)=0$ and $C_{(n-1)1}(0)=1$. } 
	\label{Fig:osc_freq_largeRabi}
\end{figure}

As an illustration of the violation of the condition for resonance separation, we show the oscillation frequencies for $|\Omega_R^{(2)}| = |\Omega_R^{(3)}| = 0.5\Omega$ in Fig.~\ref{Fig:osc_freq_largeRabi}. Here the anticrossing picture of isolated resonances is smeared and cannot be observed.

\section{Control of entangled states}

In order to control the quantum state of the system, turn the entanglement
on/off, read or write information into a qubit, or implement a logic gate,
one has to vary the parameters of a system, for example the detuning from
resonance, the field amplitude of the EM mode at the atom position, or the
intensity of a classical acoustic pumping. The analytic results obtained in
previous sections can be readily generalized when the variation of a
parameter is adiabatic, i.e. slower than the optical frequencies $\omega $
or $\frac{W}{\hbar }$. Since the space is limited, the time-dependent
problem will be considered elsewhere. Here we consider just one example,
namely turning on/off of a classical acoustic pumping $\mathbf{q}=\mathbf{Q}%
\left( \mathbf{r}\right) e^{-i\Omega t}+\mathbf{Q}^{\ast }\left( \mathbf{r}%
\right) e^{i\Omega t}.$

For maximal control it is beneficial to place an atom at the point where $%
\mathbf{E}\left( \mathbf{r=r}_{a}\right) \rightarrow 0$, whereas $\left(
\mathbf{Q\cdot \nabla }\right) \mathbf{E}_{\mathbf{r}=\mathbf{r}_{a}}$ is
maximized. The equations of motion for quantum state amplitudes were derived
in Sec.~V, see Eqs.~(\ref{cn0 dot})-(\ref{c00 dot}).

Consider an exact parametric resonance $\omega +\Omega =\frac{W}{\hbar }$
for simplicity, when%
\begin{equation*}
\omega _{n}=\omega _{n-1}+\frac{W}{\hbar }.
\end{equation*}

The solution to Eqs.~(\ref{cn0 dot})-(\ref{c00 dot}) when the acoustic
pumping is turned off is
\begin{equation*}
\Psi =C_{00}\left( 0\right) e^{-i\omega _{0}t}\left\vert 0\right\rangle
\left\vert 0\right\rangle +\sum_{n=1}^{\infty }\left( C_{n0}\left( 0\right)
e^{-i\omega _{n}t}\left\vert n\right\rangle \left\vert 0\right\rangle
+C_{\left( n-1\right) 1}\left( 0\right) e^{-i\left( \omega _{n-1}+\frac{W}{%
\hbar }\right) t}\left\vert n-1\right\rangle \left\vert 1\right\rangle
\right)
\end{equation*}

The solution when the acoustic pumping is turned on is%
\begin{eqnarray}
\Psi &=&C_{00}\left( 0\right) e^{-i\omega _{0}t}\left\vert 0\right\rangle
\left\vert 0\right\rangle +\sum_{n=1}^{\infty }[\left( A_{n}e^{-i\left\vert
\Omega _{R}^{\left( n\right) }\right\vert t}+B_{n}e^{i\left\vert \Omega
_{R}^{\left( n\right) }\right\vert t} \right) e^{-i\omega _{n}t}\left\vert
n\right\rangle \left\vert 0\right\rangle  \notag \\
&&+\left( -A_{n}e^{-i\left\vert \Omega _{R}^{\left( n\right) }\right\vert
t}+B_{n}e^{i\left\vert \Omega _{R}^{\left( n\right) }\right\vert t} \right)
e^{i\theta -i\left( \omega _{n-1}+\frac{W}{\hbar }\right) t}\left\vert
n-1\right\rangle \left\vert 1\right\rangle ]  \label{Psi}
\end{eqnarray}

Assume that the initial quantum state before the pumping was turned on was
not entangled, for example, an atom was in an excited state and there were no photons:
\begin{equation*}
\Psi =e^{-i\left( \omega _{0}+\frac{W}{\hbar }\right) t}\left\vert
0\right\rangle \left\vert 1\right\rangle. 
\end{equation*}
If the acoustic pumping is turned on at $t=0$, the quantum state becomes
entangled:%
\begin{equation}
\Psi =ie^{-i\omega _{1}t-i\theta }\sin \left( \left\vert \Omega _{R}^{\left(
1\right) }\right\vert t\right) \left\vert 1\right\rangle \left\vert
0\right\rangle +e^{-i\left( \omega _{0}+\frac{W}{\hbar }\right) t}\cos
\left( \left\vert \Omega _{R}^{\left( 1\right) }\right\vert t\right)
\left\vert 0\right\rangle \left\vert 1\right\rangle ,  \label{psi71}
\end{equation}
Then the acoustic pumping can be turned off. Depending on the turnoff moment
of time, one can obtain various entangled photon-atom states, e.g. Bell
states etc. The above reasoning is valid when the turn-on/off rate is slower
than the optical frequencies and the detuning from the two-wave resonance $
\omega =\frac{W}{\hbar }$.

\section{Conclusions} 

In conclusion, we showed how the entanglement in a system of a fermionic quantum emitter coupled to a quantized electromagnetic field in a nanocavity and quantized phonon or mechanical vibrational modes emerges in the vicinity of a parametric resonance in the system.  We developed analytic theory describing the formation and evolution of entangled quantum states, which can be applied  to a broad range of cavity quantum optomechanics problems and  emerging nanocavity strong-coupling experiments. The model includes decoherence effects due to coupling of the fermion, photon, and phonon subsystems to their dissipative reservoirs within the stochastic evolution approach, which is derived from the Heisenberg-Langevin formalism. We showed that our approach provided the results for physical observables equivalent to those obtained from the density matrix equations with the relaxation operator in Lindblad form. We derived analytic expressions for the time evolution of the quantum state and observables, and the emission spectra. The limit of a classical acoustic pumping, the control of entangled states, and the interplay between parametric and standard two-wave resonances were discussed.

\begin{acknowledgments}
This work has been supported in part by the Air Force Office for Scientific Research
through Grant No.~FA9550-17-1-0341 and by NSF Award No.~1936276. M.T. acknowledges the support from RFBR Grant No. 20-02-00100,   M.E. acknowledges the support from Federal Research Center
Institute of Applied Physics of the Russian Academy of Sciences (Project No. 0035-2019-004).

\end{acknowledgments}

\appendix

\section{The stochastic equation of evolution for the state vector}

The description of open quantum systems within the stochastic equation of
evolution for the state vector is usually formulated for a Monte-Carlo type
numerical scheme, e.g. the method of quantum jumps \cite{Scully1997,Plenio1998}. We developed an approach suitable for analytic
derivations. Our stochastic equation of evolution is basically the Schr\"{o}dinger equation modified by adding a linear relaxation operator and the
noise source term with appropriate correlation properties. The latter are
related to the parameters of the relaxation operator in such a way that the
expressions for the statistically averaged quantities satisfy certain
physically meaningful conditions.

The protocol of introducing the relaxation operator with a corresponding
noise source term to the quantum dynamics is well known in the Heisenberg
picture, where it is called the Heisenberg-Langevin method \cite{Scully1997,
Gardiner2004, Tokman2013}. We develop a conceptually similar approach for
the Schr\"{o}dinger equation. Here we derive the general form of the
stochastic equation of evolution from the Heisenberg-Langevin equations and
track how certain physically reasonable constraints on the observables
determine the correlation properties of the noise sources.

\subsection{From Heisenberg-Langevin equations to the stochastic equation
for the state vector}

The Heisenberg-Langevin equation for the operator $\widehat{g}$ of a certain
observable quantity takes the form \cite{Scully1997, Gardiner2004,
Tokman2013}

\begin{equation}
\frac{d}{dt}\hat{g}=\frac{i}{\hbar }\left[ \hat{H},\hat{g}\right] +\hat{R}%
\left( \hat{g}\right) +\hat{L}_{g}\left( t\right) ,  \label{hle}
\end{equation}%
where $\hat{R}\left( \hat{g}\right) $ is the relaxation operator, $\hat{L}%
_{g}\left( t\right) $ is the Langevin noise source satisfying $\overline{%
\hat{L}_{g}\left( t\right) }=0$, where the bar means statistical averaging.
For given commutation relations of the two operators: $\left[ \hat{g}_{1},%
\hat{g}_{2}\right] =C$, where $C$ is a constant, correct Langevin sources
should ensure the conservation of commutation relations at any moment of
time, despite the presence of the relaxation operator in Eq.~(\ref{hle});
see \cite{Tokman2013, Erukhimova2017, Tokman2019}.

The group of terms $\frac{i}{\hbar }\left[ \hat{H},\hat{g}\right] +\hat{R}%
\left( \hat{g}\right) $ can often be written as%
\begin{equation}
\frac{i}{\hbar }\left[ \hat{H},\hat{g}\right] +\hat{R}\left( \hat{g}\right) =%
\frac{i}{\hbar }\left( \hat{H}_{eff}^{\dagger }\hat{g}-\hat{g}\hat{H}%
_{eff}\right) ,  \label{effh term}
\end{equation}%
where $\hat{H}_{eff}$ is a non-Hermitian operator. For example, if the
relaxation operator describes dissipation with relaxation constant $\gamma $%
, so that $\overline{\hat{g}}\propto e^{-\gamma t}$ , then $\hat{H}_{eff}=$ $%
\hat{H}-i\hbar \frac{\gamma }{2}\hat{1}$, where $\hat{1}$ is a unit
operator. Note that in the master equation for the density matrix the
relaxation is often introduced in a conceptually similar way \cite%
{Plenio1998}, $\left[ \hat{H},\hat{\rho}\right] \Longrightarrow \hat{H}_{eff}%
\hat{\rho}-\hat{\rho}\hat{H}_{eff}^{\dagger }$, which is however slightly
different from the form used in Eq.~(\ref{effh term}): $\left[ \hat{H},\hat{g%
}\right] \Longrightarrow \hat{H}_{eff}^{\dagger }\hat{g}-\hat{g}\hat{H}%
_{eff} $. The difference is because the commutator of an unknown operator
with Hamiltonian enters with opposite sign in the master equation as
compared to the Heisenberg equation.

Now consider the transition from the Heisenberg-Langevin equation to the
stochastic equation for the state vector. The key point is to assume that
there exists the operator of evolution $\hat{U}\left( t\right) $, which is
determined not only by the system parameters but also by the properties of
the reservoir. This operator determines the evolution of the state vector:
\begin{equation}
\left\vert \Psi \left( t\right) \right\rangle =\hat{U}\left( t\right)
\left\vert \Psi _{0}\right\rangle ,\left\langle \Psi \left( t\right)
\right\vert =\left\langle \Psi _{0}\right\vert \hat{U}^{\dagger }\left(
t\right) ,  \label{evo of sv}
\end{equation}%
where $\Psi _{0}=\Psi \left( 0\right) $. Hereafter we will denote the
operators in the Schr\"{o}dinger picture with index \textquotedblleft
s\textquotedblright\ to distinguish them from the Heisenberg operators. An
observable can be calculated as
\begin{equation*}
g\left( t\right) =\left\langle \Psi \left( t\right) \right\vert \hat{g}%
_{S}\left\vert \Psi \left( t\right) \right\rangle =\left\langle \Psi
_{0}\right\vert \hat{g}\left( t\right) \left\vert \Psi _{0}\right\rangle
\end{equation*}%
Which leads to%
\begin{equation}
\hat{g}\left( t\right) =\hat{U}^{\dagger }\left( t\right) \hat{g}_{S}\hat{U}%
\left( t\right) ,  \label{g hat t}
\end{equation}%
Since the substitution of Eqs.~(\ref{evo of sv}) and~(\ref{g hat t}) into
the standard Heisenberg equation leads to the standard Schr\"{o}dinger
equation, it makes sense to apply the same procedure to the
Heisenberg-Langevin equation in order to obtain the \textquotedblleft
stochastic variant\textquotedblright\ of the Schr\"{o}dinger equation. The
solution of the latter should yield the expression for an observable,%
\begin{equation*}
g\left( t\right) =\overline{\left\langle \Psi \left( t\right) \right\vert
\hat{g}_{S}\left\vert \Psi \left( t\right) \right\rangle },
\end{equation*}%
Which is different from the standard expression by additional averaging over
the noise statistics.

Note that an open system interacting with a reservoir is generally in a
mixed state and should be described by the density matrix. We are describing
the state of the system with a state vector which has a fluctuating
component. For example, in a certain basis $\left\vert \alpha \right\rangle $
the state vector will be $C_{\alpha }\left( t\right) =\overline{C_{\alpha }}+%
\widetilde{C_{\alpha }}$, where the fluctuating component is denoted with a
wavy bar. The elements of the density matrix of the corresponding mixed
state are $\rho _{\alpha \beta }=\overline{C_{\alpha }C_{\beta }^{\ast }}=%
\overline{C_{\alpha }}\cdot \overline{C_{\beta }^{\ast }}+\overline{%
\widetilde{C_{\alpha }}\cdot \widetilde{C_{\beta }}^{\ast }}$.

The solution to the Heisenberg-Langevin equation can be expressed through
the evolution operator $\hat{U}\left( t\right) $ using Eq.~(\ref{g hat t}).
The noise source terms should be chosen to ensure the conservation of
commutation relations at any moment of time, despite the presence of the
relaxation operator. Since commutation relations between any two operators
are conserved if and only if the evolution operator $\hat{U}\left( t\right) $
is unitary, a correct noise source in the Heisenberg-Langevin equation will
automatically ensure the condition $\hat{U}^{\dagger }\hat{U}=\hat{1}$.

We implement the above protocol. Substituting Eq.~(\ref{g hat t}) together
with $\hat{H}_{eff}=\hat{U}^{\dagger }\hat{H}_{eff,S}\hat{U}$ and $\hat{H}%
_{eff}^{\dagger }=\hat{U}^{\dagger }\hat{H}_{eff,S}^{\dagger }\hat{U}$ into
Eqs.~(\ref{hle}),(\ref{effh term}), and using $\hat{U}^{\dagger }\hat{U}=%
\hat{1}$, we arrive at 
\begin{equation}
\left( \frac{d}{dt}\hat{U}^{\dagger }-\frac{i}{\hbar }\hat{U}^{\dagger }\hat{%
H}_{eff,S}^{\dagger }\right) \hat{g}_{S}\hat{U}+\hat{U}^{\dagger }\hat{g}%
_{S}\left( \frac{d}{dt}\hat{U}+\frac{i}{\hbar }\hat{H}_{eff,S}\hat{U}\right)
=\hat{L}_{g}  \label{lg}
\end{equation}%
Next, we introduce the operator $\hat{F}$, defined by%
\begin{equation}
\hat{L}_{g}=2\hat{U}^{\dagger }\hat{g}_{S}\hat{F}  \label{dof}
\end{equation}%
For the operator $\hat{L}_{g}^{\dagger }$, Eq.~(\ref{dof}) gives $\hat{L}%
_{g}^{\dagger }=2\hat{F}^{\dagger }$ $\left( \hat{U}^{\dagger }\hat{g}%
_{S}\right) ^{\dagger }=2\hat{F}^{\dagger }\hat{g}_{S}^{\dagger }\hat{U}$.
Since $\hat{g}$ and $\hat{g}_{S}$ are Hermitian operators, $\hat{L}_{g}$ has
to be Hermitian too. (One can develop the Heisenberg-Langevin formalism for
non-Hermitian operators too, for example creation or annihilation operators,
but the derivation becomes longer.) Then the operator $\hat{L}_{g}$ can be
\textquotedblleft split\textquotedblright\ between the two terms on the
left-hand side of Eq.~(\ref{lg}) using the relationship%
\begin{equation}
\hat{L}_{g}=\hat{U}^{\dagger }\hat{g}_{S}\hat{F}+\hat{F}^{\dagger }\hat{g}%
_{S}\hat{U}  \label{lg split}
\end{equation}%
Substituting the latter into Eq.~(\ref{lg}), we obtain

\begin{equation*}
\left( \frac{d}{dt}\hat{U}^{\dagger }-\frac{i}{\hbar }\hat{U}^{\dagger }\hat{%
H}_{eff,S}^{\dagger }-\hat{F}^{\dagger }\right) \hat{g}_{S}\hat{U}+\hat{U}%
^{\dagger }\hat{g}_{S}\left( \frac{d}{dt}\hat{U}+\frac{i}{\hbar }\hat{H}%
_{eff,S}\hat{U}-\hat{F}\right) =0.
\end{equation*}%
For simplicity we will assume operator $\hat{H}_{eff}$ to be constant with
time, i.e. we won't differentiate between $\hat{H}_{eff}$ and $\hat{H}%
_{eff,S}.$

The last equation is satisfied for sure if
\begin{equation}
\frac{d}{dt}\hat{U}=-\frac{i}{\hbar }\hat{H}_{eff}\hat{U}+\hat{F},\frac{d}{dt%
}\hat{U}^{\dagger }=\frac{i}{\hbar }\hat{U}^{\dagger }\hat{H}_{eff}^{\dagger
}+\hat{F}^{\dagger }.  \label{u dot and u dagger dot}
\end{equation}%
Multiplying Eqs.~(\ref{u dot and u dagger dot}) by the initial state vector $%
\left\vert \Psi _{0}\right\rangle $ from the right and from the left, we
obtain the stochastic equation for the state vector and its Hermitian
conjugate:
\begin{equation}
\frac{d}{dt}\left\vert \Psi \right\rangle =-\frac{i}{\hbar }\hat{H}%
_{eff}\left\vert \Psi \right\rangle -\frac{i}{\hbar }\left\vert R\left(
t\right) \right\rangle  \label{stochastic eq for psi}
\end{equation}

\begin{equation}
\frac{d}{dt}\left\langle \Psi \right\vert =\frac{i}{\hbar }\left\langle \Psi
\right\vert \hat{H}_{eff}^{\dagger }+\frac{i}{\hbar }\left\langle R\left(
t\right) \right\vert  \label{hc of stochastic eq for psi}
\end{equation}%
Where we introduced the notations $i\hbar \hat{F}$ $\left\vert \Psi
_{0}\right\rangle \Longrightarrow \left\vert R\left( t\right) \right\rangle $%
, $-i\hbar \left\langle \Psi _{0}\right\vert \hat{F}^{\dagger
}\Longrightarrow \left\langle R\left( t\right) \right\vert .$ We will also
need Eqs.~(\ref{stochastic eq for psi}) and~(\ref{hc of stochastic eq for
psi}) in a particular basis $\left\vert \alpha \right\rangle $:

\begin{equation}
\frac{d}{dt}C_{\alpha }=-\frac{i}{\hbar }\sum_{\nu }\left( \hat{H}%
_{eff}\right) _{\alpha \nu }C_{\nu }-\frac{i}{\hbar }R_{\alpha },
\label{c alpha dot}
\end{equation}

\begin{equation}
\frac{d}{dt}C_{\alpha }^{\ast }=\frac{i}{\hbar }\sum_{\nu }C_{\nu }^{\ast
}\left( \hat{H}_{eff}^{\dagger }\right) _{\nu \alpha }+\frac{i}{\hbar }%
R_{\alpha }^{\ast },  \label{hc of c alpha dot}
\end{equation}%
where $R_{\alpha }=\left\langle \alpha \right. \left\vert R\right\rangle $, $%
\left( \hat{H}_{eff}\right) _{\alpha \beta }=\left\langle \alpha \right\vert
\hat{H}_{eff}\left\vert \beta \right\rangle $.

Applying the same procedure to the standard Heisenberg equation~(\ref{hei})
we obtain that in Eqs.~(\ref{stochastic eq for psi}),(\ref{hc of stochastic
eq for psi}): $\hat{H}_{eff}\equiv \hat{H}_{eff}^{\dagger }=\hat{H}$\ and $%
\left\langle R\left( t\right) \right\vert \equiv 0$, which corresponds to
the standard Schr\"{o}dinger equation and its Hermitian conjugate.

Note that intermediate relations~(\ref{u dot and u dagger dot}) for the
evolution operator and in particular operator $\hat{F}$ should not depend on
the choice of a particular physical observable $g$ in the original
Heisenberg-Langevin equation~(\ref{hle}). We assume that the Langevin
operators in the original equation do not contradict this physically
reasonable requirement.

In general, statistical properties of noise that ensure certain
physically meaningful requirements impose certain constraints on the noise
source $\left\vert R\right\rangle $ which enters the right-hand side of the
stochastic equation for the state vector. In particular, it is natural to
require that the statistically averaged quantity $\overline{\left\vert
R\right\rangle }=0$. We will also require that the noise source $\left\vert
R\right\rangle $ has the correlation properties that preserve the norm of
the state vector averaged over the reservoir statistics:
\begin{equation}
\overline{\left\langle \Psi \left( t\right) \right. \left\vert \Psi \left(
t\right) \right\rangle }=1.  \label{stat n of sv}
\end{equation}

\subsection{Noise correlator}

The solution to Eqs.~(\ref{stochastic eq for psi}) and~(\ref{hc of
stochastic eq for psi}) can be formally written as%
\begin{equation}
\left\vert \Psi \right\rangle =e^{-\frac{i}{\hbar }\hat{H}_{eff}t}\left\vert
\Psi _{0}\right\rangle -\frac{i}{\hbar }\int_{0}^{t}e^{\frac{i}{\hbar }\hat{H%
}_{eff}\left( \tau -t\right) }\left\vert R\left( \tau \right) \right\rangle
d\tau ,  \label{sol to stochastic eq for psi}
\end{equation}

\begin{equation}
\left\langle \Psi \right\vert =\left\langle \Psi _{0}\right\vert e^{\frac{i}{%
\hbar }\hat{H}_{eff}^{\dagger }t}+\frac{i}{\hbar }\int_{0}^{t}\left\langle
R\left( \tau \right) \right\vert e^{-\frac{i}{\hbar }\hat{H}_{eff}^{\dagger
}\left( \tau -t\right) }d\tau ,  \label{sol to hc of stochastic eq for psi}
\end{equation}%
In the basis $| \alpha \rangle$, Eqs.~(\ref{sol to stochastic eq for psi}),(\ref{sol to hc of
stochastic eq for psi}) can be transformed into%
\begin{equation}
C_{\alpha }=\left\langle \alpha \right\vert e^{-\frac{i}{\hbar }\hat{H}%
_{eff}t}\left\vert \Psi _{0}\right\rangle -\frac{i}{\hbar }%
\int_{0}^{t}\left\langle \alpha \right\vert e^{\frac{i}{\hbar }\hat{H}%
_{eff}\left( \tau -t\right) }\left\vert R\left( \tau \right) \right\rangle
d\tau ,  \label{c alpha}
\end{equation}

\begin{equation}
C_{\alpha }^{\ast }=\left\langle \Psi _{0}\right\vert e^{\frac{i}{\hbar }%
\hat{H}_{eff}^{\dagger }t}\left\vert \alpha \right\rangle +\frac{i}{\hbar }%
\int_{0}^{t}\left\langle R\left( \tau \right) \right\vert e^{-\frac{i}{\hbar
}\hat{H}_{eff}^{\dagger }\left( \tau -t\right) }\left\vert \alpha
\right\rangle d\tau .  \label{hc of c alpha}
\end{equation}%
In order to calculate the observables, we need to know the expressions for
the averaged dyadic combinations of the amplitudes. We can find them using
Eqs.~(\ref{c alpha dot}) and~(\ref{hc of c alpha dot}):%
\begin{eqnarray}
\frac{d}{dt}\overline{C_{\alpha }C_{\beta }^{\ast }} &=&-\frac{i}{\hbar }%
\sum_{\nu }\left( H_{\alpha \nu }^{\left( h\right) }\overline{C_{\nu
}C_{\beta }^{\ast }}-\overline{C_{\alpha }C_{\nu }^{\ast }}H_{\nu \beta
}^{\left( h\right) }\right) -\frac{i}{\hbar }\sum_{\nu }\left( H_{\alpha \nu
}^{\left( ah\right) }\overline{C_{\nu }C_{\beta }^{\ast }}+\overline{%
C_{\alpha }C_{\nu }^{\ast }}H_{\nu \beta }^{\left( ah\right) }\right)  \notag
\\
&&+\left( -\frac{i}{\hbar }\overline{C_{\beta }^{\ast }R_{\alpha }}+\frac{i}{%
\hbar }\overline{R_{\beta }^{\ast }C_{\alpha }}\right) ,  \label{adc of amp}
\end{eqnarray}%
Where we separated the Hermitian and anti-Hermitian components of the
effective Hamiltonian: $\left\langle \alpha \right\vert \hat{H}%
_{eff}\left\vert \beta \right\rangle =H_{\alpha \beta }^{\left( h\right)
}+H_{\alpha \beta }^{\left( ah\right) }$. Substituting Eqs.~(\ref{c alpha})
and~(\ref{hc of c alpha}) into the last term in Eq.~(\ref{adc of amp}), we
obtain%
\begin{eqnarray*}
-\frac{i}{\hbar }\overline{C_{\beta }^{\ast }R_{\alpha }}+\frac{i}{\hbar }%
\overline{C_{\alpha }R_{\beta }^{\ast }} &=&\frac{1}{\hbar ^{2}}\int_{-t}^{0}%
\overline{\left\langle R\left( t+\xi \right) \right\vert e^{-\frac{i}{\hbar }%
\hat{H}_{eff}^{\dagger }\xi }\left\vert \beta \right\rangle \left\langle
\alpha \right. \left\vert R\left( t\right) \right\rangle }d\xi \\
&&+\frac{1}{\hbar ^{2}}\int_{-t}^{0}\overline{\left\langle R\left( t\right)
\right. \left\vert \beta \right\rangle \left\langle \alpha \right\vert e^{%
\frac{i}{\hbar }\hat{H}_{eff}\xi }\left\vert R\left( t+\xi \right)
\right\rangle }d\xi .
\end{eqnarray*}%
To proceed further with analytical results, we need to evaluate these
integrals. The simplest situation is when the noise source terms are
delta-correlated in time (Markovian). In this case only the point $\xi =0$
contributes to the integrals. As a result, Eq.~(\ref{adc of amp})) is
transformed to
\begin{equation}
\frac{d}{dt}\overline{C_{\alpha }C_{\beta }^{\ast }}=-\frac{i}{\hbar }%
\sum_{\nu }\left( H_{\alpha \nu }^{\left( h\right) }\overline{C_{\nu
}C_{\beta }^{\ast }}-\overline{C_{\alpha }C_{\nu }^{\ast }}H_{\nu \beta
}^{\left( h\right) }\right) -\frac{i}{\hbar }\sum_{\nu }\left( H_{\alpha \nu
}^{\left( ah\right) }\overline{C_{\nu }C_{\beta }^{\ast }}+\overline{%
C_{\alpha }C_{\nu }^{\ast }}H_{\nu \beta }^{\left( ah\right) }\right)
+D_{\alpha \beta },  \label{tran adc of amp}
\end{equation}%
Where the correlator $D_{\alpha \beta }$ is defined by%
\begin{equation}
\overline{R_{\beta }^{\ast }\left( t+\xi \right) R_{\alpha }\left( t\right) }%
=\overline{R_{\beta }^{\ast }\left( t\right) R_{\alpha }\left( t+\xi \right)
}=\hbar ^{2}\delta \left( \xi \right) D_{\alpha \beta }  \label{cor d}
\end{equation}%
The time derivative of the norm of the state vector is given by%
\begin{equation}
\frac{d}{dt}\sum_{\alpha }\overline{\left\vert C_{\alpha }\right\vert ^{2}}%
=-\sum_{\alpha }\left[ \frac{i}{\hbar }\sum_{\nu }\left( H_{\alpha \nu
}^{\left( ah\right) }\overline{C_{\nu }C_{\alpha }^{\ast }}+\overline{%
C_{\alpha }C_{\nu }^{\ast }}H_{\nu \alpha }^{\left( ah\right) }\right)
-D_{\alpha \alpha }\right]  \label{dot of nor of sv}
\end{equation}%
Clearly, the components $D_{\alpha \alpha }$ of the noise correlator need to
compensate the decrease in the norm due to the anti-Hermitian component of
the effective Hamiltonian. Therefore the expressions for $H_{\alpha \beta
}^{\left( ah\right) }$and $D_{\alpha \alpha }$ have to be mutually
consistent. This is the manifestation of the fluctuation-dissipation theorem
\cite{Landau1965}.

Note that the noise correlator could depend on the averaged combinations
(e.g. dyadics) of the components of the state vector. This is because the
noise source term $\left\vert R\right\rangle $ introduced above depends on
the initial state $\left\vert \Psi _{0}\right\rangle $ and the evolution
operator $\hat{U}$, and these quantities form the state vector components at
any given time. Of course, what we call a \textquotedblleft state
vector\textquotedblright\ is the solution of the stochastic equation of
motion, which is very different from the solution of the conventional Schr%
\"{o}dinger equation for a closed system. In particular, we postulated the
existence of the evolution operator $\hat{U}$ determined not only by the
parameters of the dynamical system but also the properties of a dissipative
reservoir, although we did not specify any particular expression for $\hat{U}
$.

As an example, consider a simple diagonal anti-Hermitian operator $H_{\alpha
\nu }^{\left( ah\right) }$:

\begin{equation}
H_{\alpha \nu }^{\left( ah\right) }=-i\hbar \gamma _{\alpha }\delta _{\alpha
\nu }  \label{ah operator}
\end{equation}

And introduce the following models:

(i) Populations relax much slower than coherences (expected for condensed
matter systems). In this case we can choose $D_{\alpha \neq \beta }=0$, $%
D_{\alpha \alpha }=2\gamma _{\alpha }\overline{\left\vert C_{\alpha
}\right\vert ^{2}}$; within this model the population at each state will be
preserved.

(ii) The state $\alpha =\alpha _{down}$ has a minimal energy, while the
reservoir temperature $T=0$. In this case it is expected that all
populations approach zero in equilibrium whereas the occupation number of
the ground state approaches $1$, similar to the Weisskopf-Wigner model. The
adequate choice of correlators is $D_{\alpha \neq \beta }=0$, $D_{\alpha
\alpha }\propto \delta _{\alpha \alpha _{down}}$, $\gamma _{\alpha
_{down}}=0 $. The expression for the remaining nonzero correlator,%
\begin{equation}
D_{\alpha _{down}\alpha _{down}}=\sum_{\alpha \neq \alpha _{down}}2\gamma
_{\alpha }\overline{\left\vert C_{\alpha }\right\vert ^{2}},
\label{non 0 cor}
\end{equation}

Ensures the conservation of the norm:%
\begin{equation*}
\frac{d}{dt}\sum_{\alpha \neq \alpha _{down}}\overline{\left\vert C_{\alpha
}\right\vert ^{2}}=-\sum_{\alpha \neq \alpha _{down}}2\gamma _{\alpha }%
\overline{\left\vert C_{\alpha }\right\vert ^{2}}=-\frac{d}{dt}\overline{%
\left\vert C_{\alpha _{down}}\right\vert ^{2}}.
\end{equation*}%
This is an example of the correlator's dependence on the state vector that
we discussed before.


\subsection{Comparison with the Lindblad method}

One can choose the anti-Hermitian Hamiltonian $H_{\alpha \beta }^{\left(
ah\right) }$ and correlators $D_{\alpha \beta }$ in the stochastic equation
of motion in such a way that Eq.~(\ref{tran adc of amp}) for the dyadics $%
\overline{C_{n}C_{m}^{\ast }}$ correspond exactly to the equations for the
density matrix elements in the Lindblad approach. Indeed, the Lindblad form
of the master equation has the form \cite{Scully1997,Plenio1998}%
\begin{equation}
\frac{d}{dt}\hat{\rho}=-\frac{i}{\hbar }\left[ \hat{H},\hat{\rho}\right] +%
\hat{L}\left( \hat{\rho}\right)  \label{Lind mas eq}
\end{equation}%
where $\hat{L}\left( \hat{\rho}\right) $ is the Lindbladian:%
\begin{equation}
\hat{L}\left( \hat{\rho}\right) =-\frac{1}{2}\sum_{k}\gamma _{k}\left( \hat{l%
}_{k}^{\dagger }\hat{l}_{k}\hat{\rho}+\hat{\rho}\hat{l}_{k}^{\dagger }\hat{l}%
_{k}-2\hat{l}_{k}\hat{\rho}\hat{l}_{k}^{\dagger }\right) ,
\label{Lindbladian}
\end{equation}%
Operators $\hat{l}_{k}$ in Eq.~(\ref{Lindbladian}) and their number are
determined by the model which describes the coupling of the dynamical system
to the reservoir. The form of the relaxation operator given by Eq.~(\ref%
{Lindbladian}) preserves automatically the conservation of the trace of the
density matrix, whereas the specific choice of relaxation constants ensures
that the system approaches a proper steady state given by thermal
equilibrium or supported by an incoherent pumping.

Eq.~(\ref{Lind mas eq}) is convenient to represent in a slightly different
form:
\begin{equation}
\frac{d}{dt}\hat{\rho}=-\frac{i}{\hbar }\left( \hat{H}_{eff}\hat{\rho}-\hat{%
\rho}\hat{H}_{eff}^{\dagger }\right) +\delta \hat{L}\left( \hat{\rho}\right)
\label{dot rho}
\end{equation}%
where%
\begin{equation}
\hat{H}_{eff}=\hat{H}-i\hbar \sum_{k}\gamma _{k}\hat{l}_{k}^{\dagger }\hat{l}%
_{k},~~~\delta \hat{L}\left( \hat{\rho}\right) =\sum_{k}\gamma _{k}\hat{l}%
_{k}\hat{\rho}\hat{l}_{k}^{\dagger }.  \label{eff h and delta rho}
\end{equation}%
Writing the anti-Hermitian component of the Hamiltonian in Eqs.~(\ref{c
alpha dot}),(\ref{hc of c alpha dot}) as
\begin{equation}
H_{\alpha \beta }^{\left( ah\right) }=-i\hbar \left\langle \alpha
\right\vert \sum_{k}\gamma _{k}\hat{l}_{k}^{\dagger }\hat{l}_{k}\left\vert
\beta \right\rangle ,  \label{expre for ah operator}
\end{equation}%
and defining the corresponding correlator of the noise source as
\begin{equation}
\overline{R_{\beta }^{\ast }\left( t+\xi \right) R_{\alpha }\left( t\right) }%
=\hbar ^{2}\delta \left( \xi \right) D_{\alpha \beta },~~~~D_{\alpha \beta
}=\left\langle \alpha \right\vert \delta \hat{L}\left( \hat{\rho}\right)
\left\vert \beta \right\rangle _{\rho _{mn}=\overline{C_{n}C_{m}^{\ast }}},
\label{cor of noise source}
\end{equation}%
We obtain the solution in which averaged over noise statistics dyadics $%
\overline{C_{n}C_{m}^{\ast }}$ correspond exactly to the elements of the
density matrix within the Lindblad method.

Instead of deriving the stochastic equation of evolution of the state vector
from the Heisenberg-Langevin equations we could postulate it from the very
beginning. After that, we could justify the choice of the effective
Hamiltonian and noise correlators by ensuring that they lead to the same
observables as the solution of the density matrix equations with the
relaxation operator in Lindblad form \cite{Plenio1998,blum}. However, the
demonstration of direct connection between the stochastic equation of evolution of the state vector 
and the Heisenberg-Langevin equation provides an important physical insight.


\subsection{Relaxation rates for coupled subystems interacting with a
reservoir}

Whenever we have several coupled subsystems (such as electrons, photon
modes, and phonons in this paper), each coupled to its reservoir, the
determination of relaxation rates of the whole system becomes nontrivial.
The problem can be solved if we assume that these \textquotedblleft
partial\textquotedblright\ reservoirs are statistically independent.In this
case it is possible to add up partial Lindbladians and obtain the total
effective Hamiltonian.

Consider the Hamiltonian~(\ref{toh}) of the system formed by a two-level
electron system coupled to an EM mode field and dressed by a phonon field:
\begin{equation}
\hat{H}=\hat{H}_{em}+\hat{H}_{a}+\hat{H}_{p}+\hat{V}.  \label{total ham}
\end{equation}%
Here $\hat{H}_{em}=\frac{\hbar \omega }{2}\left( \hat{c}^{\dagger }\hat{c}+%
\hat{c}\hat{c}^{\dagger }\right) $ is the Hamiltonian for a single EM mode
field, $\ \hat{H}_{a}=W_{1}\hat{\sigma}^{\dagger }\hat{\sigma}+W_{0}\hat{%
\sigma}\hat{\sigma}^{\dagger }$ is the Hamiltonian for a two-level
\textquotedblleft atom\textquotedblright\ with energy levels $W_{0,1}$, $%
\hat{H}_{p}=\frac{\hbar \Omega }{2}\left( \hat{b}^{\dagger }\hat{b}+\hat{b}%
\hat{b}^{\dagger }\right) $ is the Hamiltonian for a phonon mode, $\hat{V}=%
\hat{V}_{1}+\hat{V}_{2}$ the interaction Hamiltonian, where $\hat{V}_{1,2}$
describe the atom-photon and atom-photon-phonon coupling, respectively:
\begin{equation*}
\hat{V}_{1}=-\left( \chi \hat{\sigma}^{\dagger }\hat{c}+\chi ^{\ast }\hat{%
\sigma}\hat{c}^{\dagger }+\chi \hat{\sigma}\hat{c}+\chi ^{\ast }\hat{\sigma}%
^{\dagger }\hat{c}^{\dagger }\right) ,
\end{equation*}%
\begin{equation*}
\hat{V}_{2}=-\left( \eta _{1}\hat{\sigma}^{\dagger }\hat{c}\hat{b}+\eta
_{1}^{\ast }\hat{\sigma}\hat{c}^{\dagger }\hat{b}^{\dagger }+\eta _{2}\hat{%
\sigma}^{\dagger }\hat{c}\hat{b}^{\dagger }+\eta _{2}^{\ast }\hat{\sigma}%
\hat{c}^{\dagger }\hat{b}+\eta _{1}\hat{\sigma}\hat{c}\hat{b}+\eta
_{1}^{\ast }\hat{\sigma}^{\dagger }\hat{c}^{\dagger }\hat{b}^{\dagger }+\eta
_{2}\hat{\sigma}\hat{c}\hat{b}^{\dagger }+\eta _{2}^{\ast }\hat{\sigma}%
^{\dagger }\hat{c}^{\dagger }\hat{b}\right) ,
\end{equation*}%
where $\chi $, $\eta _{1}$, $\eta _{2}$ are coupling constants defined
before.

Summing up the known (see e.g. \cite{Scully1997,Plenio1998}) partial Lindbladians of
two bosonic (infinite amount of energy levels) and one fermionic (two-level)
subsystems, we obtain
\begin{eqnarray}
L\left( \hat{\rho}\right) &=&-\frac{\gamma }{2}N_{1}^{T_{a}}\left( \hat{%
\sigma}\hat{\sigma}^{\dagger }\hat{\rho}+\hat{\rho}\hat{\sigma}\hat{\sigma}%
^{\dagger }-2\hat{\sigma}^{\dagger }\hat{\rho}\hat{\sigma}\right) -\frac{%
\gamma }{2}N_{0}^{T_{a}}\left( \hat{\sigma}^{\dagger }\hat{\sigma}\hat{\rho}+%
\hat{\rho}\hat{\sigma}^{\dagger }\hat{\sigma}-2\hat{\sigma}\hat{\rho}\hat{%
\sigma}^{\dagger }\right)  \notag \\
&&-\frac{\mu _{\omega }}{2}\overline{n}_{\omega }^{T_{em}}\left( \hat{c}\hat{%
c}^{\dagger }\hat{\rho}+\hat{\rho}\hat{c}^{\dagger }\hat{c}-2\hat{c}%
^{\dagger }\hat{\rho}\hat{c}\right) -\frac{\mu _{\omega }}{2}\left(
\overline{n}_{\omega }^{T_{em}}+1\right) \left( \hat{c}^{\dagger }\hat{c}%
\hat{\rho}+\hat{\rho}\hat{c}\hat{c}^{\dagger }-2\hat{c}\hat{\rho}\hat{c}%
^{\dagger }\right)  \notag \\
&&-\frac{\mu _{\Omega }}{2}\overline{n}_{\Omega }^{T_{p}}\left( \hat{b}\hat{b%
}^{\dagger }\hat{\rho}+\hat{\rho}\hat{b}^{\dagger }\hat{b}-2\hat{b}^{\dagger
}\hat{\rho}\hat{b}\right) -\frac{\mu _{\Omega }}{2}\left( \overline{n}%
_{\Omega }^{T_{p}}+1\right) \left( \hat{b}^{\dagger }\hat{b}\hat{\rho}+\hat{%
\rho}\hat{b}\hat{b}^{\dagger }-2\hat{b}\hat{\rho}\hat{b}^{\dagger }\right)
\label{lindb}
\end{eqnarray}%
where $\gamma $, $\mu _{\omega }$ and $\mu _{\Omega }$ are partial
relaxation rates of the subsystems,%
\begin{equation*}
N_{0,1}^{T_{a}}=\left( 1+e^{-\frac{W_{1}-W_{0}}{T_{a}}}\right) ^{-1}e^{-%
\frac{W_{0,1}-W_{0}}{T_{a}}},\ \overline{n}_{\omega }^{T_{em}}=\left( e^{%
\frac{\hbar \omega }{T_{em}}}-1\right) ^{-1},\ \overline{n}_{\Omega
}^{T_{p}}=\left( e^{\frac{\hbar \Omega }{Tp}}-1\right) ^{-1},
\end{equation*}%
$T_{a,em,p}$ are the temperatures of partial reservoirs. For the Lindblad
master equation in the form Eq.~(\ref{dot rho}) we get%
\begin{equation}
\hat{H}_{eff}=\hat{H}-i\hat{\Gamma},  
\label{effectiv ham}
\end{equation}%
where%
\begin{equation}
\hat{\Gamma}=\frac{\hbar }{2}\left\{ \gamma \left( N_{1}^{T_{a}}\hat{\sigma}%
\hat{\sigma}^{\dagger }+N_{0}^{T_{a}}\hat{\sigma}^{\dagger }\hat{\sigma}%
\right) +\mu _{\omega }\left[ \overline{n}_{\omega }^{T_{em}}\hat{c}\hat{c}%
^{\dagger }+\left( \overline{n}_{\omega }^{T_{em}}+1\right) \hat{c}^{\dagger
}\hat{c}\right] +\mu _{\Omega }\left[ \overline{n}_{\Omega }^{T_{p}}\hat{b}%
\hat{b}^{\dagger }+\left( \overline{n}_{\Omega }^{T_{p}}+1\right) \hat{b}%
^{\dagger }\hat{b}\right] \right\} .  \label{capital gamma}
\end{equation}%
Using the effective Hamiltonian given by Eqs.~(\ref{effectiv ham}),(\ref%
{capital gamma}), we arrive at the stochastic equation for the state vector
in the following form:
\begin{equation}
\frac{d}{dt}C_{\alpha n0}=-i\frac{W_{0}+\hbar \omega \left( n+\frac{1}{2}%
\right) +\hbar \Omega \left( \alpha +\frac{1}{2}\right) }{\hbar }C_{\alpha
n0}-\frac{i}{\hbar }\left\langle \alpha \right\vert \left\langle
n\right\vert \left\langle 0\right\vert \hat{V}\left\vert \Psi \right\rangle
-\gamma _{\alpha n0}C_{\alpha n0}-\frac{i}{\hbar }R_{\alpha n0},
\label{stochastic eq for calphan0}
\end{equation}

\begin{equation}
\frac{d}{dt}C_{\alpha n1}=-i\frac{W_{1}+\hbar \omega \left( n+\frac{1}{2}%
\right) +\hbar \Omega \left( \alpha +\frac{1}{2}\right) }{\hbar }C_{\alpha
n1}-\frac{i}{\hbar }\left\langle \alpha \right\vert \left\langle
n\right\vert \left\langle 1\right\vert \hat{V}\left\vert \Psi \right\rangle
-\gamma _{\alpha n1}C_{\alpha n1}-\frac{i}{\hbar }R_{\alpha n1},
\label{stochastic eq for calphan1}
\end{equation}%
where%
\begin{equation}
\gamma _{\alpha n0}=\frac{\gamma }{2}N_{1}^{T_{a}}+\frac{\mu _{\omega }}{2}%
\left[ \overline{n}_{\omega }^{T_{em}}\left( n+1\right) +\left( \overline{n}%
_{\omega }^{T_{em}}+1\right) n\right] +\frac{\mu _{\Omega }}{2}\left[
\overline{n}_{\Omega }^{T_{p}}\left( \alpha +1\right) +\left( \overline{n}%
_{\Omega }^{T_{p}}+1\right) \alpha \right] ,  \label{gamma alphan0}
\end{equation}

\begin{equation}
\gamma _{\alpha n1}=\frac{\gamma }{2}N_{0}^{T_{a}}+\frac{\mu _{\omega }}{2}%
\left[ \overline{n}_{\omega }^{T_{em}}\left( n+1\right) +\left( \overline{n}%
_{\omega }^{T_{em}}+1\right) n\right] +\frac{\mu _{\Omega }}{2}\left[
\overline{n}_{\Omega }^{T_{p}}\left( \alpha +1\right) +\left( \overline{n}%
_{\Omega }^{T_{p}}+1\right) \alpha \right] ,  \label{gamma alphan1}
\end{equation}%
Eqs.~(\ref{gamma alphan0}),(\ref{gamma alphan1}) determine the rules of
combining the \textquotedblleft partial\textquotedblright\ relaxation rates
for several coupled subsystems.


\end{document}